\documentclass[11pt, aps, preprint, nofootinbib,superscriptaddress,eqsecnum]{revtex4-2}
\usepackage[active]{srcltx}
\usepackage[utf8]{inputenc}
\usepackage{latexsym}
\usepackage{hyperref}
\usepackage{amsmath}
\usepackage{amsfonts}
\usepackage{graphicx}
\usepackage{slashed}
\usepackage{xcolor}
\usepackage{soul}
\usepackage{slashed}
\usepackage{braket}
\usepackage{natbib}
\usepackage[caption=false]{subfig}

\usepackage{physics}

\usepackage{MnSymbol}
\usepackage{wasysym}

\begin{document}

\title{\Large{{\sc Spontaneous Symmetry Breaking \\ and Frustrated Phases}}}

\author{Heitor Casasola}
\email{heitorcasacola@uel.br}
\affiliation{Departamento de F\'isica, Universidade Estadual de Londrina, \\86057-970, Londrina, PR, Brasil}

\author{Carlos A. Hernaski}
\email{carloshernaski@utfpr.edu.br}
\affiliation{Departamento de F\'isica, Universidade Tecnológica Federal do Paraná, 
85503-390, Pato Branco, PR, Brasil}

\author{Pedro R. S. Gomes}
\email{pedrogomes@uel.br}
\affiliation{Departamento de F\'isica, Universidade Estadual de Londrina, \\86057-970, Londrina, PR, Brasil}

\author{Paula F. Bienzobaz}
\email{paulabienzobaz@uel.br}
\affiliation{Departamento de F\'isica, Universidade Estadual de Londrina, \\86057-970, Londrina, PR, Brasil}

\begin{abstract}	

We study a system involving a single quantum degree of freedom per site of the lattice interacting with a few neighbors (up to second neighbors), with the interactions chosen as to produce frustration. At zero temperature, this system undergoes several quantum phase transitions from both gapped-to-gapless and gapless-to-gapless phases, providing a very rich phase structure with disordered, homogeneous and modulated ordered phases meeting in a quantum Lifshitz point. The gapless phases spontaneously break spatial lattice translations  as well as internal symmetries of the form $U(1)^{\mathsf{N}_c}$, where $\mathsf{N}_c$ is the number of independent pitch vectors that arise in the homogeneous and modulated ordered phases. We carry out a detailed analysis of the quantum critical behavior discussing the mechanism leading to the phase transitions. We also discuss a proper characterization of all the gapless phases as well as the nature of the Goldstone excitations. We study the behavior of the correlation functions and identify regions in the phase diagram where the system exhibits generalized symmetries, such as  polynomial shift symmetry. This type of symmetry plays an important role in the so-called fractonic phases, which is an exotic form of matter recently discovered.


\end{abstract}

\maketitle

\newpage

\section{Introduction}

Frustrated spin systems exhibit very rich physical properties \cite{Diep}. In general, frustration arises when there are competing interactions so that the lattice degrees of freedom cannot be all simultaneously accommodated in such a way as to minimize the energetic cost due to the interactions. This generally introduces a high degree of fluctuations that can drive the systems to very complex phases, for example, exhibiting modulated structures \cite{Lubensky}, or even topological order \cite{Mila,Balents}.

The case of gapless  modulated phases arising as the result of spontaneous symmetry breaking (SSB) is interesting since the Goldstone excitations reflect the symmetry-breaking pattern.  In general, both internal and spatial symmetries can be spontaneously broken in a modulated phase due to the existence of a nontrivial pitch vector ${\bf q}^c$, since it introduces preferred positions and directions in the system. The pitch vector is the Fourier mode that maximizes the Fourier transform of the interaction energy $J({\bf q})=\sum_{\bf h} J(|{\bf h}|)e^{i {\bf h}\cdot {\bf q}}$, with ${\bf h}={\bf r}-{\bf r}'$. For example, for  ferromagnetic interactions between first neighbors in a cubic lattice, the pitch vector is trivial, ${\bf q}^c=0$, but for competing ferro and anti-ferro interactions we have in general ${\bf q}^c\neq 0$.

The celebrated Goldstone theorem provides a systematic way of counting Goldstone modes in the case of SSB of {\it global internal symmetries} in {\it relativistic theories}. However, this situation covers only a class of SSB phenomena (for a recent good review about SSB see \cite{Beekman}). Lattice models in general do not exhibit relativistic invariance emerging at low energies, so that frequently we deal with SSB in a nonrelativistic setup. In addition, in a modulated phase such as envisioned above, the SSB may involve spatial symmetries. In these situations the counting of Goldstone excitations is more subtle. Although some of these cases have been studied many years ago \cite{Nielsen}, it was only recently that great progresses in the extension of the Goldstone theorem have been achieved \cite{Low,Brauner,Watanabe,Watanabe1,Hidaka,Watanabe2} (see also the recent review \cite{Watanabe3}).

The central purpose of this work is to study gapless frustrated phases that are able to support nonrelativistic Goldstone excitations. To address this question we use as a prototype a quantum system constituted of a lattice model with a single bosonic degree of freedom per site, interacting with a few neighboring sites. The interactions are set as to produce frustration. In addition, the imposition of a global constraint involving all degrees of freedom of the system is able to drive the system to both thermal and quantum phase transitions. Models of this type are known as quantum spherical models (QSM) \cite{Obermair,Henkel,Nieuwenhuizen,Vojta1,Gracia,Oliveira}. This class of models has the advantage of being exactly solvable in arbitrary dimension while exhibiting nontrivial critical behavior. It is therefore especially suitable for our purposes because it allows us to study the effects of frustration without introducing any extra complications, making it possible to proceed analytically. Naturally, we intend to further extend these studies to other systems in future investigations.

The minimum phase structure of the QSM arising when there is no competing interactions comprises a disordered phase (gapped) and an ordered one (gapless). The nature of this transition has some peculiarities, which we revisit here from the perspective of SSB. It is similar to a Bose-Einstein condensation and, as we shall discuss in detail, corresponds to a transition from a phase where the symmetry is {\it explicitly} broken to a phase where the symmetry is {\it spontaneously} broken; there is no phase where the symmetry is realized exactly (linearly). In this sense, it is surprising that this transition belongs to the universality class of the $O(N)$ nonlinear sigma model (in the large $N$ limit) \cite{Stanley,Vojta1,Gomes2}, since this latter undergoes a transition from a phase where the symmetry is exact to a phase with SSB. 

\begin{figure}[!h]
\centering
\includegraphics[scale=0.8]{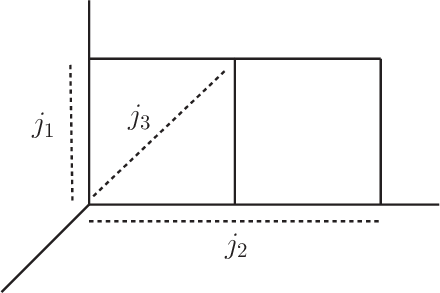}
\caption{Competing interactions in a two-dimensional slice of a hypercubic lattice.}
\label{CompeticaoFig}
\end{figure}

When frustrated interactions are introduced the phase structure of the QSM becomes much richer \cite{Nussinov,Nussinov0,Nussinov1,Paula}. We will consider a hypercubic lattice where the first-neighbor interactions are ferromagnetic, with strength $j_1$, whereas the second-neighbor and first-diagonal-neighbor interactions are anti-ferromagnetic, with strengths $j_2$ and $j_3$, respectively. This arrangement is depicted in Fig. \ref{CompeticaoFig}.

We carry out a detailed study of the quantum critical behavior in $d=2$ and $d=3$ spatial dimensions. Beyond the gapped-to-gapless phase transition mentioned above (that takes place even without competition, i.e., $j_2=j_3=0$), we have identified several distinct regions in the parameter space according to the relative strength of the interactions $j_1,j_2$ and $j_3$, corresponding to different homogeneous and modulated ordered phases. We show that these ordered phases are separated by gapless-to-gapless continuum quantum phase transitions, that manifest as discontinuities in the critical chemical potential\footnote{The chemical potential is introduced in this model as a Lagrange multiplier enforcing the mean spherical constraint (see eq. \ref{mean}).} of the model as we cross such regions. 

We discuss in detail the characterization of the ordered phases associated with each region of the parameter space through the identification of the pitch vectors and the construction of the corresponding ordered ground states. This enables us to determine the symmetries that are spontaneously broken. Once the interaction preserves certain discrete spatial symmetries, we have a certain number $\mathsf{N}_c$ of critical modes ${\bf q}^c$. In a generic ordered phase the internal symmetry group that is spontaneously broken is given by $U(1)^{\mathsf{N}_c}$. Furthermore the subgroup of lattice translations that preserves the set of critical modes defines a residual translation subgroup of the phase. In many modulated phases there is no translation invariance at all. 

In this scenario, we analyze how to define excitations above the ordered ground states and we perform the counting of independent Goldstone modes due to the spontaneous breaking of the global continuous symmetries $U(1)^{\mathsf{N}_c}$. In spite of the number of broken charges being greater than one for the modulated phases, we show that there is always only a single Goldstone excitation because all the charges have the same local generator, i.e., locally they are indistinguishable (non-uniform symmetry). We have also identified certain special regions in the phase diagrams where the system exhibits exotic behavior, associated with the presence of an enhanced symmetry, known in the literature as polynomial shift symmetries \cite{Horava0,Horava1,Horava2,Gromov}.

This work is organized as follows. In Sec. (\ref{Sec2}) we discuss some general properties of the model that only depend on simple assumptions about the interaction function. We show that as in the minimal QSM we should have a gapped-to-gapless phase transition and that the competing ferro and anti-ferro interactions can give rise to homogeneous and modulated gapless phases, each one associated with a specific set of pitch vectors. In Sec. (\ref{sec3}), we make a qualitative analysis of the gapped-to-gapless phase transition using mean field approximation. In Sec. (\ref{sec4}) we discuss a specific form of the interaction function that presents the desired properties discussed in the previous sections. We then calculate the pitch vectors that arise in the ordered phases for $d=2$ and $d=3$, and delimit the regions in parameter space of each ordered phase. Sec. (\ref{sec5}) is devoted to an analysis of the critical behavior of the model focusing on the analytic properties of the free energy. We show that besides the gapped-to-gapless continuous phase transition already present in the minimal QSM, we also find continuous gapless-to-gapless transitions between the ordered phases specified by distinct classes of pitch vectors. In Sec. (\ref{sec6}) we discuss some details of the ordered phases, as the explicit construction of the ground states and the global symmetries that are spontaneously broken in each phase. We also perform in this section the counting of Goldstone excitations and show that due to the non-uniform character of the symmetries only one Goldstone mode can be excited for any one of the phases. We address further discussions in Sec. (\ref{sec7}).


\section{Quantum Spherical Spins: Spectrum}\label{Sec2}

The quantum spherical model (QSM) is defined by the following Hamiltonian
\begin{equation}
\mathcal{H}= 
\frac{g}{2}\sum_{\bf{r}}\Pi_{\bf{r}}^2-\frac{1}{2}\sum_{\bf{r},{\bf{r}'}}J_{\bf{r},{\bf{r}'}} S_{\bf{r}}S_{\bf{r}'},
\label{1.1}
\end{equation}
with the spins subject to the global constraint
\begin{equation}
\sum_{\bf{r}}S_{\bf{r}}^2=N
\label{1.2}
\end{equation}
and satisfying the canonical quantization relation
\begin{equation}
[S_{\bf{r}},\Pi_{\bf{r}'}]=i\delta_{\bf{r},{\bf{r}'}}.
\label{1.3}
\end{equation}
The parameter $g$ characterizes the quantum fluctuations and plays a similar role as the temperature in the case of thermal fluctuations, providing the distance to the (quantum) critical point. The interaction between neighbors is parameterized by $J_{\bf{r},{\bf{r}'}}$, which we assume to depend only on the distance between the sites, i.e., $J_{\bf{r},{\bf{r}'}}\equiv J(|\bf{r}-\bf{r}'|)$. Adopting periodic boundary conditions, we get a lattice translation invariance. Therefore, denoting ${\bf r}=\left(x_1,\ldots,x_d\right)$, we can immediately verify the invariance of the model under the following spatial discrete transformations:

i) {\it Lattice translations}
\begin{equation}
x_i~~\rightarrow~~ x_i+a_i,~~~\text{with}~~~a_i~\in~\mathbb{Z}.
\label{1.31} 
\end{equation}

ii) {\it Permutations}
\begin{equation}
\left(x_1,\ldots,x_d\right)~~\rightarrow~~ \mathcal{P}_{i_1,\ldots,i_d}\left(x_1,\ldots,x_d\right)=\left(x_{i_1},\ldots,x_{i_d}\right).
\label{1.32}
\end{equation}
The set of $\mathcal{P}_{i_1,\ldots,i_d}$, with $i_1,\ldots,i_d=1,\ldots,d$ and $i_j\neq i_k$ forms the group of permutations of the components $x_i$. For example, $\mathcal{P}_{2,3,1}\left(x_1,x_2,x_3\right)=\left(x_2,x_3,x_1\right)$.

iii) {\it Signal change of components}
\begin{equation}
\left(x_1,\ldots,x_d\right)~~\rightarrow~~ \mathcal{S}_{n_1,\ldots,n_d}\left(x_1,\ldots,x_d\right)=\left(\left(-1\right)^{n_1}x_1,\ldots,\left(-1\right)^{n_d}x_d\right),
\label{1.33}
\end{equation}
with $n_i=0,1$.
For example, $\mathcal{S}_{0,1,1}\left(x_1,x_2,x_3\right)=\left(x_1,-x_2,-x_3\right)$.

The QSM is exactly solvable in the thermodynamic limit even in the presence of linearly coupled external fields. The simplest way to determine the spectrum of the model and to unveil its phase structure is to consider the global spherical constraint in average, which is a thermodynamically equivalent formulation. So, we just add a term $\frac{\mu}{2} \sum_{\bf r}S_{\bf r}^2$ in the Hamiltonian, with $\mu$ being a variational constant parameter that will be used to implement the mean spherical constraint. In this case the constraint (\ref{1.2}) is replaced by
\begin{equation}
\langle \sum_{\bf r}S_{\bf r}^2\rangle = -\frac{2}{\beta} \frac{\partial}{\partial \mu} \ln Z \equiv N,
\label{mean}
\end{equation}
with $Z$ being the partition function. The Hamiltonian in this formulation describes a set of coupled harmonic oscillators. Therefore, it is convenient to use Fourier analysis to bring the Hamiltonian to a diagonal form. Let us consider the Fourier decomposition
\begin{equation}
S_{\bf r}=\frac{1}{\sqrt{N}}\sum_{\bf q} e^{i {\bf q}\cdot {\bf r}}S_{\bf q}~~~\text{and}~~~\Pi_{\bf r}=\frac{1}{\sqrt{N}}\sum_{\bf q} e^{i {\bf q}\cdot {\bf r}}\Pi_{\bf q},
\label{1.4}
\end{equation}
with the modes restricted to the first Brillouin zone,
\begin{equation}
-\pi<q_i\leq \pi. \label{1.41}
\end{equation}
With this, the Hamiltonian becomes
\begin{eqnarray}
\mathcal{H}=\frac{g}{2}\sum_{\bf{q}}\Pi_{\bf{q}}\Pi_{-\bf{q}}+\frac{1}{2g}\sum_{\bf{q}}\omega_{\bf q}^2 S_{\bf{q}} S_{-\bf{q}},
\label{1.5}
\end{eqnarray}
where we have defined the frequency
\begin{equation}
\omega_{\bf q}^2\equiv g \left(\mu-J({\bf q})\right)
\label{1.6}
\end{equation}
and the Fourier transform of the interaction energy
\begin{equation}
J({\bf q})=\sum_{\bf h}J(|{\bf h}|)e^{i {\bf q}\cdot{\bf h}},~~~\text{with}~~~{\bf h}={\bf r}-{\bf r}^{\prime}.
\label{1.7}
\end{equation}
From (\ref{1.32}) and (\ref{1.33}), it follows that J({\bf q}) is symmetric under the analogue operations in momentum space
\begin{eqnarray}
\left(q_1,\ldots,q_d\right)~~&\rightarrow&~~ \mathcal{P}_{i_1,\ldots,i_d}\left(q_1,\ldots,q_d\right);\label{1.62a}\\
\left(q_1,\ldots,q_d\right)~~&\rightarrow&~~ \mathcal{S}_{n_1,\ldots,n_d}\left(q_1,\ldots,q_d\right).\label{1.62}
\end{eqnarray}
Let us consider modes ${\bf q}$ and ${\bf q}^\prime$ that are related by the transformations (\ref{1.62a}) and (\ref{1.62}), i.e., ${\bf q}'= \mathcal{P}\mathcal{S}({\bf q})$. We will denote by $\left[{\bf q}\right]$ the set of all momenta connected to a given ${\bf q}$ in this way. Due to these properties we have that $\omega_{{\bf q}\prime}=\omega_{{\bf q}}$ if ${\bf q}'= \mathcal{P}\mathcal{S}({\bf q})$.

The form of the frequency (\ref{1.6}) implies the stability condition 
\begin{equation}
\mu \geq \text{max}\,J({\bf q}).
\label{1.8}
\end{equation}
To illustrate the role of the critical modes (pitch vectors), we consider a simple scenario where the maximum of $J({\bf q})$ occurs for some isolated  critical mode family $\left[{\bf q}^c\right]$, that is, $\text{max}\,J({\bf q})\equiv J\left(\left[{\bf q}^c\right]\right)\equiv \mu_c$. Let us denote by $\mathsf{N}_c$ the number of modes in the set $[{\bf q}^c]$. We notice from (\ref{1.6}) that the frequency of the critical mode family vanishes as $\mu\rightarrow\mu_c$. Therefore, it is convenient to consider the two cases: $\mu=\mu_c$ and $\mu>\mu_c$.

For $\mu>\mu_c$, since always $\omega_{\bf q}\neq 0$, we can define the usual annihilation and creation operators
\begin{eqnarray}
a_{\bf q}&\equiv&\left(\frac{\omega_{\bf q}}{2g}\right)^{\frac{1}{2}}S_{\bf q}+i \left(\frac{g}{2\omega_{\bf q}}\right)^{\frac{1}{2}}\Pi_{\bf q},\label{1.9}\\
a^\dagger_{\bf q}&\equiv&\left(\frac{\omega_{\bf q}}{2g}\right)^{\frac{1}{2}}S^\dagger_{\bf q}-i \left(\frac{g}{2\omega_{\bf q}}\right)^{\frac{1}{2}}\Pi^\dagger_{\bf q},\label{1.10}
\end{eqnarray}
and bring the Hamiltonian to the familiar form  
\begin{equation}
\mathcal{H}=\sum_{\bf q} \omega_{\bf q}\left(a_{\bf q}^{\dagger}a_{\bf q}+\frac{1}{2}\right).
\label{1.11}
\end{equation}
For this range of values of the parameter $\mu$, the energy spectrum is simply
\begin{equation} 
E_{n_{\bf q}}=\sum_{{\bf q}}\omega_{\bf q}\left(n_{\bf q}+\frac{1}{2}\right),~~~~~\text{with}~~n_{\bf q}=0,1,2,\ldots\,.\label{1.12}
\end{equation}
Therefore, the model presents a unique vacuum, $n_{{\bf q}}=0$, which is the unique state annihilated by the $a_{\bf q}$ operators.

For $\mu=\mu_c$, we can use the relations (\ref{1.9}) and (\ref{1.10}) only for the non-critical modes. For the case that ${\bf q}^c=0$, we have only one critical mode, $\mathsf{N}_c=1$, since $\mathcal{P}\mathcal{S}{\left(0,\ldots,0\right)}=\left(0,\ldots,0\right)$ and the Hamiltonian becomes
\begin{equation}
\mathcal{H}=\frac{g}{2}\left(P_{0}\right)^2+\sum_{{\bf q}\neq 0} \omega_{\bf q}\left(a_{\bf q}^{\dagger}a_{\bf q}+\frac{1}{2}\right),
\label{1.13}
\end{equation}
with $P_0\equiv\Pi_0$ being a Hermitian conserved charge.
For ${\bf q}^c=\left(n_1,\ldots,n_d\right)\pi$ with $n_i=0,1$, we use (\ref{1.41}) to get $\mathcal{S}\left(n_1\pi,\ldots,n_d\pi\right)=\left(n_1\pi,\ldots,n_d\pi\right)$. If some $n_i\neq 0$, $\mathcal{P}({\bf q}^c)$ generates other nontrivial members in the set $\left[{\bf q}^c\right]$. In this case, the Hamiltonian will be of the type
\begin{equation}
\mathcal{H}=\frac{g}{2}\sum_{{\bf q}\,\in\, [{\bf q}^c] }P_{{\bf q}}^2+\sum_{{\bf q}\,\notin\,\left[{\bf q}^c\right]} \omega_{\bf q}\left(a_{\bf q}^{\dagger}a_{\bf q}+\frac{1}{2}\right),
\label{1.13b}
\end{equation}
where for each ${\bf q}^c=(n_1,\ldots,n_d)\pi$ we have defined
\begin{equation}
P_{\left(n_1\pi,\ldots,n_d\pi\right)}\equiv \Pi_{\left(n_1\pi,\ldots,n_d\pi\right)}=\sum_{{\bf r}}\Pi_{{\bf r}}\left(-1\right)^{n_1x_1}\ldots\left(-1\right)^{n_dx_d},
\label{1.13c}
\end{equation}
which are Hermitian conserved charges in the model.

For more general ${\bf q}=\left(q_1,\ldots,q_d\right)\,\in [{\bf q}^c]$ having at least one $q_j$ such that $q_j\neq n\pi$, the Fourier components $\Pi_{{\bf q}}$ will not be Hermitian. Since $\Pi_{{\bf q}}\Pi_{-{\bf q}}=\Pi_{{\bf q}}\Pi^\dagger_{{\bf q}}=\left(\left|\text{Re}\Pi_{{\bf q}}\right|^2+\left|\text{Im}\Pi_{{\bf q}}\right|^2\right)$, it is convenient to define the Hermitian operators $P^{(1)}_{{\bf q}}\equiv\text{Re}\Pi_{{\bf q}}$ and $P^{(2)}_{{\bf q}}\equiv\text{Im}\Pi_{{\bf q}}$ and rewrite the Hamiltonian in the form
\begin{equation}
\mathcal{H}=\frac{g}{2}\sum_{{\bf q}\,\in\,[{\bf q}^c]}\left(\left(P^{(1)}_{{\bf q}}\right)^2+\left(P^{(2)}_{{\bf q}}\right)^2\right)+
\sum_{{\bf q}\,\notin\,\left[{\bf q}^c\right]} \omega_{\bf q}\left(a_{\bf q}^{\dagger}a_{\bf q}+\frac{1}{2}\right).
\label{1.14}
\end{equation}
We also define the operators $X^{(1)}_{{\bf q}}\equiv\text{Re}S_{{\bf q}}$ and $X^{(2)}_{{\bf q}}\equiv\text{Im}S_{{\bf q}}$. The $P$-$X$ operators obey the commutation rules
\begin{eqnarray}
\left[X^{(k)}_{{\bf q}}, X^{(l)}_{{\bf q}^\prime}\right] =\left[P^{(k)}_{{\bf q}}, P^{(l)}_{{\bf q}^\prime}\right]=0~~~\text{and}~~~
\left[X^{(k)}_{{\bf q}}, P^{(l)}_{{\bf q}^\prime}\right] =\frac{i\delta_{kl}}{2}\left(\delta_{{\bf q},{\bf q}^\prime}-\left(-1\right)^k\delta_{{\bf q},-{\bf q}^\prime}\right),
\label{1.18a}
\end{eqnarray}
with $k,l=1,2$, since according to our definitions 
\begin{equation}
X_{-{\bf q}}^{(k)}=(-1)^{k+1}X_{\bf q}^{(k)}~~~\text{and}~~~ P_{-{\bf q}}^{(k)}=(-1)^{k+1}P_{\bf q}^{(k)}.
\label{1.18b}
\end{equation}

For any class $\left[{\bf q}^c\right]$ discussed above, the Hamiltonians correspond to a sum of free particles plus harmonic oscillators. Since the spin variables are limited, $-\sqrt{ N}\leq\text{Re} S_{{\bf q}}\leq \sqrt{ N}$ and $-\sqrt{ N}\leq\text{Im} S_{{\bf q}}\leq \sqrt{ N}$, the spectrum of the momenta operators are 
\begin{equation}
P^{(k)}_{{\bf q}}~\rightarrow ~\frac{\pi n^{(k)}_{\bf q}}{\sqrt{ N}},~~~\text{with}~~~n_{-{\bf q}}^{(k)}=(-1)^{k+1}n_{\bf q}^{(k)}, ~~~n^{(k)}_{\bf q} \in \mathbb{Z}.
\label{1.15}
\end{equation} 
The energy spectrum is then
\begin{eqnarray}
E^N_{n^{(i)}_{\bf q},n_{\bf q}}&=& \frac{g \pi^2}{2 N}\sum_{{\bf q}\,\in\,[{\bf q}^c]}\sum_{i=1}^{2}\left(n^{(i)}_{\bf q}\right)^2+\sum_{{\bf q}\,\notin\,\left[{\bf q}^c\right]} \omega_{\bf q}\left(n_{\bf q}+\frac{1}{2}\right)\nonumber\\
&=&\frac{g \pi^2}{ N}\sum_{{\bf q}\,\in\,[{\bf q}^c]/\mathbb{Z}_2}\sum_{i=1}^{2}\left(n^{(i)}_{\bf q}\right)^2+\sum_{{\bf q}\,\notin\,\left[{\bf q}^c\right]} \omega_{\bf q}\left(n_{\bf q}+\frac{1}{2}\right), ~~~n_{\bf q}=0,1,2,\ldots.
\label{1.16}
\end{eqnarray}
where the sum can be restricted to the coset $[{\bf q}^c]/\mathbb{Z}_2$ (defined by the equivalence relation ${\bf q}\sim -{\bf q}$), because of the constraints in (\ref{1.15}). The first term on the right-hand side corresponds to the Anderson tower of states \cite{Beekman}, whose appearance is a striking feature of SSB mechanism. These states possess a gap of order $\sim 1/N$ and consequently vanish in the thermodynamic limit, making the ground state infinitely degenerated, 
\begin{equation}
E^{\infty}_{n^{(i)}_{\bf q},0}=\frac{1}{2}\sum_{{\bf q}\,\notin\,\left[{\bf q}^c\right]} \omega_{\bf q}, 
\label{1.17}
\end{equation}
for any $n^{(i)}_{\bf q}$, allowing for spontaneous symmetry breaking. This scenario will be analyzed in more detail in Sec. \ref{sec6}.

From (\ref{1.12}), the partition function for $\mu>\mu_c$ is that of a set of decoupled harmonic oscillators. For $\mu=\mu_c$, we have
\begin{eqnarray}
Z&=&\prod_{\bf q\,\in\, [{\bf q}^c]/\mathbb{Z}_2}\prod_{i=1}^{2}\sum_{n^{(i)}_{\bf q}=-\infty}^{\infty}\exp\left( -\frac{ \beta g \pi^2 \left(n^{(i)}_{\bf q}\right)^2}{ N}\right)\,\times\, \prod_{{\bf q}\,\notin\, \left[{\bf q}^c\right]}\left[\sum_{n_{\bf q}=0}^{\infty} \exp\left( -\beta\omega_{\bf q} \big{(}n_{{\bf q}}+\frac{1}{2}\big{)}\right) \right]\nonumber\\
&=&\left[\vartheta_3\left(0,e^{-\frac{ \beta g \pi^2}{ N}}\right)\right]^{\mathsf{N}_c}\,\times\, \prod_{{\bf q}\,\notin\, \left[{\bf q}^c\right]}\left[2 \text{sinh}\left(\frac{\beta \omega_{\bf q}}{2}\right)\right]^{-1},
\label{1.18}
\end{eqnarray}
where $\vartheta_3$ is the elliptic theta function, whose asymptotic behavior for large $N$ is
\begin{equation}
\vartheta_3\left(0,e^{-\frac{ \beta g \pi^2}{ N}}\right)~\sim~ \sqrt{\frac{ N }{\pi \beta g}}, ~~~\text{for}~N\rightarrow \infty.
\label{1.19}
\end{equation} 
In this way, the free energy is
\begin{eqnarray}
f&=&-\frac{1}{\beta N} \ln Z\nonumber\\
&=&-\frac{\mathsf{N}_c}{\beta N}  \ln \left(\frac{ N }{\pi \beta g}\right)+\frac{1}{\beta N}\sum_{{\bf q}\,\notin\, \left[{\bf q}^c\right]}\ln\left[ 2 \text{sinh}\left(\frac{\beta \omega_{\bf q}}{2}\right)\right].
\label{1.20}
\end{eqnarray}
We see that the contribution of the tower of states disappears in the thermodynamic limit and does not affect thermodynamic quantities.


\section{Qualitative analysis of the phase structure \label{sec3}}

From our previous discussion we expect a transition from a gapped phase when $\mu>\mu_c$ to a gapless phase when $\mu=\mu_c$. Besides, to satisfy the mean spherical constraint (\ref{mean}) the chemical potential gets dependent on the $g$ and $J_{{\bf r},{\bf r}^\prime}$ parameters of the Hamiltonian. Therefore, to proceed with the analysis of the quantum critical behavior of the spherical model, we need to specify the form of the interactions. Nevertheless, even with partial information about $J_{{\bf r},{\bf r}^\prime}$, we can extract interesting information from the phase transition that spring up through the dependence of $\mu$ with $g$ that are not related to the competition between the interaction parameters. In this section, we use a sort of mean field approximation to get a qualitative picture of this phase transition. As we shall argue, some of the results are exact in spite of the approximations and others will be refined in a more detailed discussions in later sections. 

From the free energy, we can derive the mean spherical constraint (\ref{mean}) at zero temperature simply by taking the limit
\begin{equation}
\langle \sum_{\bf r}S_{\bf r}^2 \rangle=\lim_{\beta\rightarrow\infty}-\frac{2}{\beta}\frac{\partial}{\partial\mu}\ln Z \equiv N.
\label{2.1}
\end{equation}
This leads to the condition
\begin{equation}
1=\frac{1}{N}\sum_{\bf q} \frac{g}{2\sqrt{g(\mu-J(\bf q))}},
\label{2.2}
\end{equation}
which determines the parameter $\mu$. This mean condition is equivalent to the strict constraint in the thermodynamic limit and for this reason does not affect thermodynamic properties of the model.

As already noticed before, we shall have
\begin{equation}
\mu-\max J({\bf q})>0.
\label{2.3}
\end{equation}
Expanding $J({\bf q})$ around the critical momentum, we get
\begin{equation}
J({\bf q})=\mu_c-O({\bf q}-{\bf q}^c),~~~\text{with}~~~O({\bf q}-{\bf q}^c)\geq0.
\label{2.3b}
\end{equation}
Thus we can write (\ref{2.2}) as
\begin{equation}
1=\frac{1}{N}\sum_{\bf q} \frac{g}{2\sqrt{g(\mu-\mu_c+O({\bf q}-{\bf q}^c)))}}.
\label{2.4}
\end{equation}
When $\mu>\mu_c$, the above sum has no singularities, even for the mode ${\bf q}={\bf q}^c$, so that it gives a well defined relation between $g$ and $\mu$. As $\omega_{\bf q}$ never vanishes in this range, this opens a gap in the system corresponding to a disordered phase\footnote{We are implicitly assuming that all the integrations over momentum are convergent. Otherwise, the system does not exhibit a phase transition.}.

In the case $\mu=\mu_c$, on the other hand, we need to be careful because there is a potential singularity in the sum. Actually, there is an indeterminacy when we go to the thermodynamic limit. To see this, we extract from the sum the contribution due to the critical mode\footnote{For simplicity, we are assuming here that there is a single critical mode, but as we have seen, this is not always the case.},
\begin{equation}
\frac{2}{\sqrt{g}}=\frac{1}{N\sqrt{\delta\mu}}+\frac{1}{N}\sum_{{\bf q}\,\notin\, \left[{\bf q}^c\right]}\frac{1}{\sqrt{\delta\mu+O({\bf q}-{\bf q}^c))}},
\label{2.5}
\end{equation} 
where $\delta\mu\equiv \mu-\mu_c$. We see that the first term on the right-hand side presents an indeterminacy in the thermodynamic limit since we have the product $N\sqrt{\delta\mu}$ with $N\rightarrow\infty$ and $\delta\mu\rightarrow 0$. 

At this point, we approximate the second term on the right-hand side as 
\begin{equation}
\frac{1}{N}\underbrace{\sum_{{\bf q}\,\notin\, \left[{\bf q}^c\right]} \frac{1}{\sqrt{\delta\mu+O({\bf q}-{\bf q}^c))}}}_{\text{Sum of order}~ N}~\rightarrow~\frac{1}{N}\left[ \frac{N}{\sqrt{\delta\mu+\bar{O}}}\right],
\label{2.6}
\end{equation}
where $\bar{O}$ corresponds to the value of $O({\bf q}-{\bf q}^c)$ taken at certain mode that supposedly provides the best approximation to the left-hand side. This leads to a tremendous simplification in the analysis, but it still captures the essence of the mechanism of phase transition concerning the variation of the $g$-parameter. With this, the complicated form (\ref{2.5}) reduces to a simple algebraic equation,
\begin{equation}
\frac{2}{\sqrt{g}}\sim\frac{1}{N\sqrt{\delta\mu}}+ \frac{1}{\sqrt{\delta\mu+\bar{O}}},
\label{2.7}
\end{equation} 
giving 
\begin{equation}
\delta\mu\left(\delta\mu+\bar{O}-\frac{g}{4}\right)=O(1/N)\rightarrow 0.
\label{2.8}
\end{equation}
We have then the two solutions in the thermodynamic limit
\begin{equation}
\mu=\mu_c~~~\text{and}~~~\mu=\mu_c + \frac{g-g_c}{4},
\label{2.9}
\end{equation}
where we have identified the critical value of the parameter $g$ as $g_c\equiv4\,\bar{O}$\footnote{It is worth to mention that in the cases we shall discuss later on, we have in general a critical surface $g_c$, spanned by varying the interaction parameters $J_{{\bf r},{\bf r}'}$. The present analysis focus on an arbitrary point on this surface.}. Recalling the condition that $\mu\geq \mu_c$, we then have the following situation:
\begin{equation}
\mu =
\begin{cases}
\;\mu_c, &\!  ~ g\leq g_c \\
\;\mu_c+\frac{g-g_c}{4}, &\!  ~ g>g_c
\end{cases}.
\label{2.10}
\end{equation}
As the relation between ($\mu-\mu_c$) and ($g-g_c$) does not depend on dimension, this result turns out to be accurate only above the upper critical dimension (mean-field). In particular, (\ref{2.10}) implies that the first order derivative $\frac{d\mu}{dg}$ is discontinuous at $g_c$.

To summarize, expressing the frequency as $\omega_{\bf q}^2\sim \mu-\mu_c+O({\bf q}-{\bf q}^c)$, we see that in the case $g>g_c$, where $\mu>\mu_c$, the frequency $\omega_{\bf q}$ never vanishes and opens a gap in the system corresponding to a disordered phase. On the other hand, for $g\leq g_c$, $\mu$ is fixed at the critical value $\mu=\mu_c$. In this case, the system is gapless, signaling a spontaneously ordered phase.

To further understand the nature of the ordered phase, it is instructive to consider the spherical constraint written in terms of momentum modes and isolating the critical mode,
\begin{equation}
\langle S_{{\bf q}^c}^{\dagger} S_{{\bf q}^c}\rangle+\sum_{{\bf q}\,\notin\, \left[{\bf q}^c\right]}\langle S_{\bf q}^{\dagger} S_{\bf q}\rangle=N.
\label{2.11}
\end{equation}
Now, at zero temperature, the expectation value at a particular mode ${\bf q}$ reads
\begin{equation}
\langle S_{{\bf q}}^{\dagger} S_{{\bf q}}\rangle=\frac{g}{2\omega_{\bf q}},
\label{2.12}
\end{equation}
which leads us back, of course, to the constraint equation (\ref{2.2}). However, these expressions are instructive since they enable us to interpret $\langle S_{{\bf q}}^{\dagger} S_{{\bf q}}\rangle$ as the ``occupation" number of the mode ${\bf q}$ and to understand the behavior of the occupation number of the critical mode as we cross the critical point $g_c$.

The mechanism of phase transition is quite similar to the Bose-Einstein condensation. Above the critical value $g_c$, we see that all modes are approximately equally occupied, which means that $\langle S_{\bf q}^{\dagger} S_{\bf q}\rangle\sim O(1)$ for all modes, including the critical mode, in order to satisfy the constraint. This is the ``normal" phase. Below the critical point, $g\leq g_c$, the critical mode is macroscopically occupied so that $\langle S_{{\bf q}^c}^{\dagger} S_{{\bf q}^c}\rangle$ becomes comparable to $N$. This corresponds to the ``condensate" phase. To quantify the fraction of $N$ that occupies the critical mode, i.e., the fraction of $N$ that is in the condensate phase, we define the order parameter $m$:
\begin{equation}
\langle S_{{\bf q}^c}^{\dagger} S_{{\bf q}^c}\rangle\equiv N m^2.
\label{2.13}
\end{equation}

In the next section we will discuss in more detail the spontaneous symmetry breaking mechanism of the model and we will explicitly construct the ground state of the ordered phase. This will confirm that $m$ indeed plays the role of an order parameter and is related to the expectation value of $S_{\bf r}$. In this way, relation (\ref{2.11}) yields to 
\begin{equation}
N m^2+\frac{\sqrt{g}}{2}\sum_{{\bf q}\,\notin\, \left[{\bf q}^c\right]} \frac{1}{\sqrt{\delta\mu+O({\bf q}-{\bf q}^c))}}=N.
\label{2.14}
\end{equation}
Using again the approximation (\ref{2.6}), this expression turns into
\begin{equation}
N m^2+\frac{\sqrt{g}}{2} \frac{N}{\sqrt{\delta\mu+\bar{O}}}=N.
\label{2.15}
\end{equation}  
Below the critical point we have $\delta\mu=0$, and consequently
\begin{equation}
m = \left(\frac{\sqrt{g}_c-\sqrt{g}}{\sqrt{g}_c}\right)^{\frac12},
\label{2.16}
\end{equation}
from which we can read the critical exponent $\beta=\frac12$. It is important to emphasize that although we have obtained this critical exponent within an approximation, it turns out to be a robust property of the spherical model in the sense that it is independent of the details of the interactions and of the dimension. The only hypothesis is that we are in large enough spatial dimensions (above the lower critical dimension) so that the system is able to order, namely, in dimensions where there is no infrared divergences. If this condition is met, then the order parameter always vanishes as we reach the critical point from below according to (\ref{2.16}).


\section{Frustrated Interactions\label{sec4}}

In the previous section we have analyzed some general properties of the quantum phase transition in the spherical model. The interactions were assumed to depend on the lattice coordinates through $|{\bf r}-{\bf r}'|$, which implies spatially homogeneity and certain discrete symmetries. In addition, we assumed that the interaction function in Fourier space is limited and its extrema can occur either at zero or non-zero momenta. One possibility to realize this situation is to consider competing ferro and anti-ferromagnetic interactions. This feature of the interaction leads to the phenomenon of frustration where the conflicting sub-interactions favor different ordering,  giving rise to a rich phase structure. In this section we provide an explicit form for the microscopic interactions that can present this kind of competition and we investigate finer details of the phase structure that depend on the specific form of the interaction.

To be explicit, we consider the following Hamiltonian:
\begin{equation}
\mathcal{H}=\frac{g}{2}\sum_{\bf r}\Pi_{\bf r}^2-j_1\sum_{<{\bf r},{\bf r}'>}S_{\bf r}S_{{\bf r}'}+j_2\sum_{\ll{\bf r},{\bf r}'\gg}S_{\bf r}S_{{\bf r}'}+
j_3\sum_{\prec{\bf r},{\bf r}'\succ}S_{\bf r}S_{{\bf r}'},
\label{4.24}
\end{equation}
with $j_1>0$ favoring ferromagnetic ordering, and $j_2>0$ and $j_3>0$ favoring anti-ferromagnetic ones. In our notation, $<>$ and $\ll\gg$ represent sums restricted to first and second neighbors of a hypercubic lattice, respectively, whereas $\prec\,\succ$ means a sum restricted to first diagonal neighbors. A two-dimensional slice is depicted in Fig. \ref{CompeticaoFig}.


\subsection{Two-Dimensional Lattice \label{IVA}}

The frustration effect in the lattice is dictated by the relative strength of $j_1,j_2$, and $j_3$. To examine this, we consider the Fourier transform of the interaction energy,
\begin{equation}
J({\bf q})=\sum_{\bf h}J(|{\bf h}|)e^{i {\bf q}\cdot{\bf h}},~~~\text{with}~~~{\bf h}={\bf r}-{\bf r}^{\prime}.
\label{0.25}
\end{equation}

It is instructive to study firstly the two-dimensional lattice, where $J({\bf q})$ becomes
\begin{equation}
J({\bf q})=2j_1[\cos(q_x)+\cos(q_y)]-2j_2[\cos(2q_x)+\cos(2q_y)]-4j_3\cos(q_x)\cos(q_y).
\label{0.26}
\end{equation}
To characterize the relative strength of the interactions, we introduce the positive parameter
\begin{equation}
p\equiv \frac{(4j_2+2j_3)}{j_1}.
\label{0.27}
\end{equation}
The dominance of ferromagnetic interactions corresponds to small values of $p$, while for large values of $p$ the anti-ferro interactions dominate. Then, at some intermediate value of $p$ a phase transition between distinct ordered phases must take place (for fixed $g<g_{c}$) . In addition, even in the region of large $p$ the two anti-ferro interactions compete and further phase transitions take place.

The maxima of $J(\bf q)$ depend on $j_1,j_2$, and $j_3$, corresponding to some set of critical modes $\left[{\bf q}^c\right]$, as we have discussed in Sec. (\ref{Sec2}). The  space of parameters can be partitioned into disjoint regions, each one governed by a specific set of critical momenta. For $d=2$, the sets $\left[{\bf q}^c\right]$ are given by
\begin{equation}
\begin{cases}
\; \left(0,0\right) &\!  \text{for}~~~p<1~~~\text{or for}~~~p=1~\text{and}~j_2\neq 0~~(I) \\
\; \left[\left(0,q\right)\right]~~~~\text{with}~~-\pi<q\leq \pi &\!  \text{for}~~~p=1~~~\text{and}~~~j_2=0~~(\ast)\\
\; \left[\left(\cos^{-1}\left(\frac{1}{p}\right),\cos^{-1}\left(\frac{1}{p}\right)\right)\right] &\!  \text{for}~~~p\geq 1~~~\text{and}~~~j_3\leq 2j_2~~(II)\\
\; \left[(q_x,q_y)\right]~~~\text{with}~~\displaystyle\sum_{i=1}^2\cos\left(q_i\right)=\frac{2}{p} &\!  \text{for}~~~p\geq 1~~~\text{and}~~~j_3=2j_2~~(II\bigcap III)\\
\; \left[\left(0,\cos^{-1}\left(\frac{j_1-2j_3}{4j_2}\right)\right)\right] &\!  \text{for}~~~p\geq 1,~~j_3\geq 2j_2,~~\text{and}~~~j_1\geq -4j_2+2j_3~~(III)\\
\; \left[\left(0,\pi\right)\right] &\!  \text{for}~~~p>1,~~j_3>2j_2,~~\text{and}~~~j_1\leq -4j_2+2j_3~~(IV).
\end{cases}
\label{0.28}
\end{equation}

Generically, for a given set of $j$-parameters, the system will be found in one of the regions displayed above, which defines a unique set $\left[{\bf q}^c(j)\right]$, except for $p=1$ with $j_2=0$ and for $p>1$ with $j_3=2j_2$, where the sets $\left[{\bf q}^c\right]$ is not uniquely fixed by the $j$-parameters. Therefore, given the parameters $j_1$, $j_2$, and $j_3$, the system will be in a disordered phase for $g>g_c$. As $g$ is lowered down to $g\leq g_c$, the system goes to an ordered or modulated phase characterized by a given set $\left[{\bf q}^c\right]$ according to the region that the $j$-parameters belong. 

As we will discuss more generally, for any spatial dimension $d$ and for $g\leq g_c$ the value $p=1$ defines a three-dimensional surface in parameter space that separates the homogeneous ordered phase from some of the possible modulated ones.  Specifically for $d=2$ and at the value $p=1$, we can show that $g_c=0$, and the surface becomes two-dimensional. The value $g_c=0$ means that the system cannot order when is on this surface due to strong quantum fluctuations. This can also be understood from the following argument. At low energies we can consider the expansion of $J(\bf {q})$ around the respective  ${\bf q}^c$. On the surface  $p=1$, we can expand around ${\bf q}^c=(0,0)$ to obtain:
\begin{eqnarray}
J(\delta{\bf q})&=&4(j_1-j_2-j_3)-(j_1-4j_2-2j_3)(\delta{\bf q})^2\nonumber\\&+&\frac{1}{4!}(2j_1-32j_2-4j_3)(\delta q_x^4+\delta q_y^4)-j_3 \delta q_x^2\delta q_y^2+\cdots\,.
\label{0.31}
\end{eqnarray}
Then, by making $p=1$ yields to $j_1-4j_2-2j_3=0$, and all the second derivatives of $J({\bf q})$ vanish. The fourth order terms then become relevant, and in the gapless phase the system is invariant under the {\it anisotropic} scaling $t \rightarrow l^2 t$ and ${\bf r}\rightarrow l {\bf r}$, with the dynamical critical exponent $z=2$. This low-energy behavior plagues the equation (\ref{2.4}) with infrared (IR) divergences and no finite value of $g$ can be found to satisfy the constraint. This result also reflects the fact that for $z=2$ the lower critical dimension for phase transition is $d=2$, according to the proper nonrelativistic version of the Mermim-Wagner theorem \cite{Horava1}. The surface defined by $g=g_c$ and $p=1$ corresponds to a  meeting of disordered, ordered, and modulated phases. The points on this surface are called quantum Lifshitz points \cite{Ardonne}.

In the next section we will make a detailed investigation of the transitions between the regions in (\ref{0.28}). In particular, we will argue that we can identify aspects of further transitions between the ordered phases (gapless-to-gapless phase transitions).


\subsection{Higher Dimensional Lattice \label{sec4b}}

The generalization of the previous discussions to an arbitrary $d$-dimensional lattice is immediate and we will be brief here.

The Fourier transform of the interaction energy is
\begin{eqnarray}
J({\bf q})=2 j_1\sum_{i=1}^d\cos q_i-2j_2\sum_{i=1}^d\cos 2q_i
- 4j_3\sum_{i<j}^d\cos q_i\cos q_j,
\label{0.35}
\end{eqnarray}
and the parameter $p$ introduced in (\ref{0.27}) is generalized to 
\begin{equation}
p\equiv \frac{[4j_2+2(d-1)j_3]}{j_1}.
\label{0.36}
\end{equation}
For $p\leq 1$ and for $p>1$ with $j_3\leq 2j_2$, we still have an analogous situation to $d=2$:
\begin{equation}
\begin{cases}
\; \left(0,\dots,0\right),  &\!   \text{for}~~~p<1~~~\text{or for}~~~p=1~\text{and}~j_2\neq 0~~(I)\\
\; \left[\left(0,\ldots,q_i,0,\ldots,0\right)\right] ~i=1,...,d &\!  \text{for}~~~p = 1~~~\text{and}~~~j_2= 0~~(\ast)\\
\; \left[\left(\cos^{-1}\left(\frac{1}{p}\right),\ldots,\cos^{-1}\left(\frac{1}{p}\right)\right)\right], ~i=1,...,d &\!  \text{for}~~~p\geq 1~~~\text{and}~~~j_3\leq 2j_2~~(II)\\
\; \left[\left(q_1,\ldots,q_d\right)\right]~~\text{with}~~\displaystyle\sum_{i=1}^d\cos{q_i}=\frac{d}{\tilde{p}} &\!  \text{for}~~~p\geq 1~~~\text{and}~~~j_3=2j_2~~(II\bigcap III).
\end{cases}
\label{0.37}
\end{equation}
However in the region $p>1$ with $j_3>2j_2$ the possible critical momenta become exceedingly richer as the dimension increases. We list the possibilities for $d=3$:
\begin{equation}
p>1~~\text{and}~~j_3>2j_2~\Rightarrow~\begin{cases}
\; \left[\left(0,\cos^{-1}\left(\frac{j_1}{4j_2}\right),\pi\right)\right] &\!  \text{for}~~~j_1\leq 4j_2~~(III)\\
\; \left[\left(0,0,\pi\right)\right] &\!  \text{for}~~~4j_2\leq j_1\leq 4j_3-4j_2~~(IV)\\
\; \left[\left(0,0,\cos^{-1}\left(\frac{j_1-4j_3}{4j_2}\right)\right)\right] &\!  \text{for}~~~j_1\geq 4j_3-4j_2~~(V).
\end{cases}
\label{0.38}
\end{equation}

We can also obtain interesting models by further tuning the interaction parameters. Exotic models emerge at low energies at the line $p=1$. Then, to investigate this possibility we expand $J({\bf q})$ around ${\bf q}^c=(0,0,\ldots,0)$,
\begin{eqnarray}
J(\delta{\bf q})&=& 2d[j_1-j_2-(d-1)j_3]-[j_1-4j_2-2(d-1)j_3]|\delta{\bf q}|^2\nonumber\\&+&\frac{1}{12}[j_1-16j_2-2(d-1)j_3]\sum_i^d \delta q_i^4-j_3\sum_{i<j}\delta q_i^2\delta q_j^2+\cdots
\label{0.39}
\end{eqnarray}
and tune the parameters to the line $p=1$ by taking $j_1-4j_2-2(d-1)j_3=0$ in this expression. If we then take $j_3=2j_2$, we obtain an emergent rotational symmetry,
\begin{equation}
J(\delta{\bf q})=2d(1+2d)j_2-j_2|\delta{\bf q}|^4+\cdots\,.
\label{0.40}
\end{equation}
On the other hand, if we adjust the parameters to $p=1$ and then take $j_2=0$, we get
\begin{equation}
J(\delta{\bf q})=2d(d-1)j_3-j_3\sum_{i<j}\delta q_i^2\delta q_j^2+\cdots\,,
\label{0.41}
\end{equation}
which, in the gapless phase, yields to the dispersion relation $E^2\sim \sum_{i<j}\delta q_i^2\delta q_j^2$. Therefore, there is an infinite number of modes with zero energy. If we set, say, $\delta q_1\neq 0$, and all the other $\delta q_j=0$, we still get zero energy. That is a kind of UV/IR mixing, in the sense that even modes near the ends of the Brillouin zone correspond to zero energy. Since the system is able to order in $d\geq 3$ dimensions, this implies an infinite number of zero modes and, as we shall discuss, an infinite number of spontaneously broken charges. This UV/IR phenomenon gives rise to remarkable physical properties and has also been observed in certain exotic type of phases of matter exhibiting fractonic behavior \cite{Fisher,Seiberg1,Seiberg2,Seiberg3}.

The existence of an infinite number of broken charges is the consequence of a large amount of symmetries arising as we take $p=1$ with $j_2=0$.  To understand this point, let us define the plaquette operator 
\begin{equation}
\Delta_{ij}S\equiv S_{(\ldots,x_i+1,\ldots,x_j+1,\ldots)}-S_{(\ldots,x_i+1,\ldots,x_j,\ldots)}-S_{(\ldots,x_i,\ldots,x_j+1,\ldots)}+S_{(\ldots,x_i,\ldots,x_j,\ldots)}.
\label{0.41a}
\end{equation}
Then we see that the interaction terms involving first and diagonal neighbors in the Hamiltonian (\ref{4.24}) reduce to
\begin{equation}
\sim\sum_{i<j} (\Delta_{ij}S)^2.
\label{0.41b}
\end{equation}
This structure is invariant under the shifts
\begin{equation}
S_{\bf r}\rightarrow S_{\bf r}+ \sum_{k=1}^d f_{k}(x_k),
\label{0.41c}
\end{equation}
with $f_{k}(x_k)$ being an arbitrary function depending only on the coordinate $x_{k}$\footnote{A two-dimensional lattice model involving this type of symmetry was studied in \cite{Fisher} in the case of {\it compact} bosons.}. In the gapless phase, this is the symmetry underlying the large number of broken charges. We shall return to this point in Sec. \ref{sec6}.


\section{Quantum Critical behavior\label{sec5}}

From the discussions of the previous sections we have gathered enough information to make a more detailed analysis of the type of phase transitions that can occur in the model. The nontrivial aspects of the spherical model dwell in the gap equation (\ref{2.2}). That equation together with the explicit form of $J({\bf q})$ will be the main focus of analysis in this section. For simplicity, we call $j_1$, $j_2$ and $j_3$ simply by $j$ when there is no need to specify which $j_i$ we are referring to or when we refer to all three $j_i$ simultaneously.


To begin with, let us reconsider the gap equation
\begin{equation}
g^{-1/2}=\frac{1}{N}\sum_{\bf q} \frac{1}{2\sqrt{\mu-J({\bf q},j)}}.
\label{4.1}
\end{equation}
Since $J({\bf q},j)$ is a smooth function of ${\bf q}$ and $j$, the integrand is also a positive smooth function of ${\bf q}$ and $j$'s, except when $\mu=\mu_c(j)$ and ${\bf q}$ $\in$ $\left[{\bf q_c}(j)\right]$. Then, for finite $N$ the sum defines a function $g=f_N(\mu,j)$ that is smooth for $\mu>\mu_c(j)$ and singular for $\mu=\mu_c(j)$. When we take the thermodynamic limit $N\rightarrow\infty$, the sum becomes an integral evaluated within the first Brillouin zone
\begin{equation}
g^{-1/2}=\int\frac{d^d{\bf q}}{2\sqrt{\mu-J({\bf q},j)}}.
\label{4.2}
\end{equation}
Except when $p=1$, $J({\bf q},j)$ will be dominated by quadratic terms in ${\bf q}$. Then, by a dimensional analysis we conclude that this integral is convergent in the IR for any dimension $d\geq 2$.

A rough analysis indicates that in the thermodynamic limit,
\begin{equation}
g(\mu,j)=\lim_{N\rightarrow\infty} f_N(\mu,j)\equiv f(\mu,j)
\end{equation}
is a smooth function of the variables for $\mu>\mu_c$, and the gap equation prescribes a unique value of $g$ for each $\mu$ and $j$ down to the value $g_c$ defined by the limit
\begin{equation}
g_c(j)=\lim_{\mu\rightarrow\mu_c(j)}f(\mu,j).
\label{4.3}
\end{equation}
However, as discussed in Sec. (\ref{sec3}), a careful analysis reveals that we can find other solutions to the gap equation, extending the range of values of $g$ down to zero. In fact, it may happen that the zero mode attains macroscopic population when the chemical potential deviation $\delta\mu\equiv\mu-\mu_c$ scales as $1/N^2$ for large $N$. 

To account for this phenomenon, let us separate the zero mode in the sum (\ref{4.1}), before taking the thermodynamic limit:
\begin{equation}
\frac{1}{\sqrt{g}}=\frac{\mathsf{N}_c}{2N\sqrt{\mu-\mu_c(j)}}+\frac{1}{N}\sum_{{\bf q}\,\notin\, \left[{\bf q}^c\right]}\frac{1}{2\sqrt{(\mu-J(\bf q))}}.
\label{4.3b}
\end{equation}
For large $N$ the second term can be approximated by the convergent integral (\ref{4.2}). Then, for $g>g_c$ we have finite $\delta\mu$, so that the first term goes to zero and we recover the solution $f(\mu,j)$. On the other hand, we can find a solution for $g<g_c$ if $\delta\mu$ scales as $\sim 1/N^2$. Since the second term tends to $\frac{1}{\sqrt{g_c}}$ as $\mu\rightarrow\mu_c$, we have the solution
\begin{equation}
\delta\mu=\frac{\mathsf{N}^2_c \,g\, g_c}{4N^2\left(\sqrt{g_c}-\sqrt{g}\right)^2},~~~~\text{for}~~~~g < g_c.
\label{4.4}
\end{equation}
Making $N\rightarrow\infty$ we get again $\mu=\mu_c(j)$ for $g < g_c$.

Thus, we can gather all the analysis as follows:
\begin{eqnarray}
\mu(g,j)=
\begin{cases}
\;\mu_c(j) &\!  ~ g\leq g_c(j) \\
\;f^{-1}(g,j) &\!  ~ g\geq g_c(j)\;.
\end{cases}
\label{4.5a}
\end{eqnarray}

\subsection{$g$-driven Phase Transition}

In Sec. (\ref{sec3}), we derived the behavior of the chemical potential $\mu$ as function of $g$ when we cross the surface $g_c$ with fixed $j$. There, we obtained a gapped-to-gapless phase transition with a discontinuity in the first derivative of $\mu$ with respect $g$. This amounts to a discontinuity in the second derivative of the free energy, characterizing a continuous phase transition. We shall verify that the mean-field result is only accurate for $d>3$, since for $d\leq 3$ the quantum fluctuations become stronger.

Since $f(\mu,j)$ is a smooth and monotonically increasing function of $\mu$ for $\mu>\mu_c$, from the inverse function theorem, $\mu(g,j)=f^{-1}(g,j)$ is smooth for $g>g_c$. In this case, to study the behavior of $\mu$ as a function of $g$, we can use the property
\begin{eqnarray}
\left(\frac{\partial\mu}{\partial g}\right)_j&=&1/\left(\frac{\partial g}{\partial \mu}\right)_j,~~~g>g_c.
\label{4.5}
\end{eqnarray}
From (\ref{4.2}), it follows that
\begin{equation}
\left(\frac{\partial g}{\partial \mu}\right)_j=g^{3/2}\int\frac{d^d{\bf q}}{2\left(\mu-J(\bf q)\right)^{3/2}}>0,~~~\mu>\mu_c.
\label{4.6}
\end{equation}
For $\mu\rightarrow\mu^+_c$ this integral only converges in the IR for $d> 3$. Then, for $d>3$  we have
\begin{equation}
\left(\frac{\partial \mu}{\partial g}\right)_j=
\begin{cases}
\; 0 &\!~ g\rightarrow g^{-}_c\\
\; \neq 0 &\! ~ g\rightarrow g^{+}_c\;
\end{cases},
\label{4.7}
\end{equation}
which shows that $\left(\frac{\partial \mu}{\partial g}\right)_j$ is discontinuous at $g=g_c$ for $d>3$.

For $d\leq 3$, the integral in (\ref{4.6}) diverges and from (\ref{4.5a}) and (\ref{4.5}) we see that $\left(\frac{\partial \mu}{\partial g}\right)_{g_{c}^{-}}=\left(\frac{\partial \mu}{\partial g}\right)_{g_{c}^{+}}=0$. Then, we investigate the second derivative in the region $g>g_{c}$, where we have the identity
\begin{eqnarray}
\left(\frac{\partial^2\mu}{\partial g^2 }\right)_j&=&-\frac{1}{\left(\frac{\partial g}{\partial \mu}\right)^3_j}\left(\frac{\partial^2 g}{\partial \mu^2}\right)_j, ~~~~~ g>g_{c}.
\label{4.8}
\end{eqnarray}
From (\ref{4.6}), we obtain
\begin{equation}
\left(\frac{\partial^2 g}{\partial\mu^2}\right)_j=\frac{3}{2}\left[g^{-1}\left(\frac{\partial g}{\partial \mu}\right)^2-g^{3/2}\int\frac{d^d{\bf q}}{2\left(\mu-J(\bf q)\right)^{5/2}}\right],~~~~~\mu>\mu_c.
\label{4.9}
\end{equation}
On dimensional grounds, we verify that $\frac{\partial g}{\partial \mu}$ is linearly divergent, and the integral in the second term has cubic divergence in the IR. Therefore, the expression (\ref{4.8}) has a finite limit as $\mu\rightarrow\mu^{+}_c$ for $d=2$ and diverges for $d=3$. Taking into account that $\mu(g,j)= \mu_{c}(j)$ for $g<g_{c}$, we then have a discontinuity in the second derivative of $\mu$:
\begin{equation}
\left(\frac{\partial^2 \mu}{\partial g^2}\right)_j=
\begin{cases}
\; 0 &\!~ g\rightarrow g^{-}_c\\
\; \neq 0 &\! ~ g\rightarrow g^{+}_c\;
\end{cases}.
\label{4.10}
\end{equation}
In this way we conclude that for both $d=2$ and $3$ the transition from the disordered phase, $g>g_c$, to the ordered phase, $g\leq g_c$, occurs closing the gap $\delta \mu$ with a third-order discontinuity in free energy.


\subsection{$j$-driven Phase Transitions}

Let us turn our attention to the possible phase transitions that occur when we vary the $j$'s maintaining $g$ fixed.


\subsubsection{Gapped-to-Gapless Transition ($g<g_c \leftrightarrow g>g_c$)}

Since $g_c$ is a function of $j$, when $j$ is varied for a fixed $g$ one can cross from a region with $g_c(j)<g$, where the system is disordered, to a region with $g_c(j)>g$, where the system is ordered. Without loss of generality, let us consider some specific value $j^{\ast}$ such that, for a small neighborhood, when $j>j^{\ast}$ we have $g_c(j)<g(\mu,j)$, and for $j<j^{\ast}$ we have $g_c(j)>g(\mu,j)$. To exemplify this situation, we have marked in Figure \ref{alldiagrams2d} \subref{diagram2d-(c)} an arbitrary $j^{\ast}_3$ in a curve $\sqrt{g}~vs~j_3$ for $d=2$ that satisfies these conditions. Let us investigate the possible source of discontinuities in this transition, i.e, how smooth is $\mu(g,j)$ when we vary $j$, with $g$ fixed, crossing the surface $g_c(j)$.

It is convenient to define the function $\mathcal{O}\left({\bf q},j\right)\equiv \mu_c(j)-J({\bf q},j)$ and write $g^{-1/2}$ as
\begin{equation}
g^{-1/2}=\int\frac{d^d{\bf q}}{2\left(\mu-\mu_c(j)+\mathcal{O}\left({\bf q},j\right)\right)^{1/2}}.\label{4.11}
\end{equation}
By definition $\mu_c(j)\equiv J({\bf q}_c,j)$, which implies $\mathcal{O}\left({\bf q}_c,j\right)=0$. Furthermore, expanding $\mathcal{O}\left({\bf q},j\right)$ around ${\bf q}_c$ gives the IR behavior $\mathcal{O}\left({\bf q},j\right)\sim \sum_{ij}\lambda_{ij}(j)\delta q_i\delta q_j$, with $\lambda_{ij}(j)$ being a positive definite matrix. 

To study the behavior of $\mu$ with respect to $j$ (with fixed $g$), it is helpful the property 
\begin{equation}
\left(\frac{\partial\mu}{\partial j}\right)_g=-\left(\frac{\partial g}{\partial j}\right)_\mu/\left(\frac{\partial g}{\partial \mu}\right)_j,~~~g>g_c.
\end{equation}
Using this relation together with (\ref{4.6}), it follows that
\begin{eqnarray}
\left(\frac{\partial \mu}{\partial j}\right)_g&=&-\left(g^{3/2}\int \frac{d^d{\bf q}}{2\left(\mu-\mu_c+\mathcal{O}\left({\bf q},j\right)\right)^{3/2}}\right)^{-1}\nonumber\\
&\times&g^{3/2}\int \frac{d^d{\bf q}}{2\left(\mu-\mu_c+\mathcal{O}\left({\bf q},j\right)\right)^{3/2}}\left(-\frac{d\mu_c}{dj}+\frac{\partial\mathcal{O}}{\partial j}\right).\label{4.12a}
\end{eqnarray}
For $j\rightarrow j^{\ast +}$, we get
\begin{equation}
\left(\frac{\partial \mu}{\partial j}\right)\bigg|_{j\rightarrow j^{\ast +}}=\left(\frac{d\mu_c}{dj}\right)\bigg|_{j\rightarrow j^{\ast +}}-\left[\left(\int \frac{d^d{\bf q}}{2\left(\mathcal{O}\left({\bf q},j\right)\right)^{3/2}}\right)^{-1}\int \frac{d^d{\bf q}}{2\left(\mathcal{O}\left({\bf q},j\right)\right)^{3/2}}\frac{\partial\mathcal{O}}{\partial j}\right]\bigg|_{j\rightarrow j^{\ast +}}.
\label{4.12f}
\end{equation}
On the other hand, for $j\rightarrow j^{\ast -}$ we have $g<g_c$ and $\mu(g,j)=\mu_c(j)$. Therefore,
\begin{equation}
\left(\frac{\partial \mu}{\partial j}\right)\bigg|_{j\rightarrow j^{\ast -}}=\left(\frac{d\mu_c}{dj}\right)\bigg|_{j\rightarrow j^{\ast -}}.\label{4.12g}
\end{equation}

For $d>3$ both integrals in the second term of (\ref{4.12f}) are convergent and, for generic points where $\frac{\partial\mathcal{O}}{\partial j}\neq 0$, we get a finite contribution from the second term. Furthermore, as we will explicit show in the next section, $\frac{d\mu_c}{dj}$ can only have discontinuities at the intersection between the regions. Since $j^{\ast}$ is not at the intersection, $\left(\frac{d\mu_{c}}{d j}\right)$ is continuous at $j^{*}$, and comparing  (\ref{4.12f})  with (\ref{4.12g}) we get a discontinuity in the first derivative $\left(\frac{\partial\mu}{\partial j}\right)\Big|_{j^{\ast}}$.

For $d\leq 3$ we can proceed similarly. The integral in the denominator of (\ref{4.12f}) diverges whereas the integral in the numerator converges. The conclusion is that $\left(\frac{\partial\mu}{\partial j}\right)_g$ is continuous at $j^{\ast}$.  Nevertheless, we can find a discontinuity at higher order derivatives. Indeed, by taking the derivative of (\ref{4.12a}) and making $j\rightarrow j^{\ast +}$, we get
\begin{eqnarray}
\left(\frac{\partial^2 \mu}{\partial j^2}\right)\bigg|_{j\rightarrow j^{\ast +}}&=&\frac{d^2\mu_c}{dj^2}\bigg|_{j\rightarrow j^{\ast +}}+\\
&+& \left[\frac{3}{2}\left(\int \frac{d^d{\bf q}}{2\left(\mathcal{O}\left({\bf q},j\right)\right)^{3/2}}\right)^{-3}\int\frac{d^d{\bf q}}{2\left(\mathcal{O}\left({\bf q},j\right)\right)^{5/2}}\left(\int \frac{d^d{\bf q}}{2\left(\mathcal{O}\left({\bf q},j\right)\right)^{3/2}}\frac{\partial\mathcal{O}}{\partial j}\right)^{2}\right]\bigg|_{j\rightarrow j^{\ast +}}.\label{4.12b} \nonumber 
\end{eqnarray}
On dimensional grounds, we conclude that the second term converges for $d=2$ and diverges for $d=3$. Thus, for generic points where $\frac{\partial\mathcal{O}}{\partial j}\neq 0$, this characterizes a discontinuity in the second derivative $\left(\frac{\partial^2 \mu}{\partial j^2}\right)_g$ when crossing the surface $g_c(j)$.

Therefore, the final conclusion is that whenever we cross the critical surface by varying either $g$ or $j$, we obtain the same kind of discontinuity in the chemical potential, characterizing a gapped-to-gapless phase transition.


\subsubsection{Gapless-to-Gapless Transitions ($g<g_c$)}

We have seen in Sec. (\ref{sec4}) that given a fixed $g\leq g_c$ we can cross ordered regions with different $[{\bf q}^c]$ families. It is not clear at this point if the crossings between these regions are fully continuous or they can contain further discontinuities in the free energy, which would signalize further phase transitions in the model. Then, let us now discuss the structure of phase diagrams when we vary $j$ with fixed $g$ but remaining always in the region $g<g_c(j)$. In this case the space dimension and the details of the interactions seem to play a major role. We concentrate in the two cases: $d=2$ and $d=3$.

\vspace{.3cm}
\underline{$d=2$}
\vspace{.3cm}

For $d=2$, when $g\leq g_c(j)$, the table (\ref{0.28}) shows the possible values of ${\bf q}^c$ and the regions in the $j$-parameter space where the function $J({\bf q})$ develops a maximum. Then, it is natural to ask if upon varying the $j$'s these phases are just continuously interconnected or further phase transitions can occur. 

Since for $g\leq g_c(j)$ we have $\mu=\mu_c(j)$, we can address this question by analyzing the analytic properties of $\mu_c(j)$. A simple calculation of the maximum of $J({\bf q})$ for the ${\bf q}^c$ values shown in the table (\ref{0.28}), gives
\begin{equation}
\mu_c(j)=
\begin{cases}
\; 4(j_1-j_2-j_3) &\!   \text{for}~~~p<1~~~\text{or for}~~~p=1~\text{and}~j_2\neq 0~~~(I) \\
\; 2j_1~~~~\text{with}~~ &\!  \text{for}~~~p=1~~~\text{and}~~~j_2=0~~~(\ast) \\
\; 4j_2 + \frac{j_1^2}{2j_2 + j_3} &\!  \text{for}~~~p\geq 1~~~\text{and}~~~j_3\leq 2j_2~~~(II)\\
\; \frac{j_1^2}{4j_2} + 4j_2~~ &\!  \text{for}~~~p\geq 1~~~\text{and}~~~j_3=2j_2~~~(II\bigcap III)\\
\; \frac{8j_1j_2+\left(j_1-2j_3\right)^2}{4j_2} &\!  \text{for}~~~p\geq 1,~~j_3\geq 2j_2,~~\text{and}~~~j_1\geq -4j_2+2j_3~~~(III)\\
\; 4(j_3-j_2) &\!  \text{for}~~~p>1,~~j_3>2j_2,~~\text{and}~~~j_1\leq -4j_2+2j_3~~~(IV).
\end{cases}
\label{4.12}
\end{equation}
As discussed in Sec. \ref{IVA}, the system is not able to order in the region $(\ast)$ indicated above, so that we not consider it in the analysis for $d=2$. However, for $d=3$ this case is relevant and it will be treated later in this section.

\begin{figure}[!h]
	\centering
	\includegraphics[scale=1]{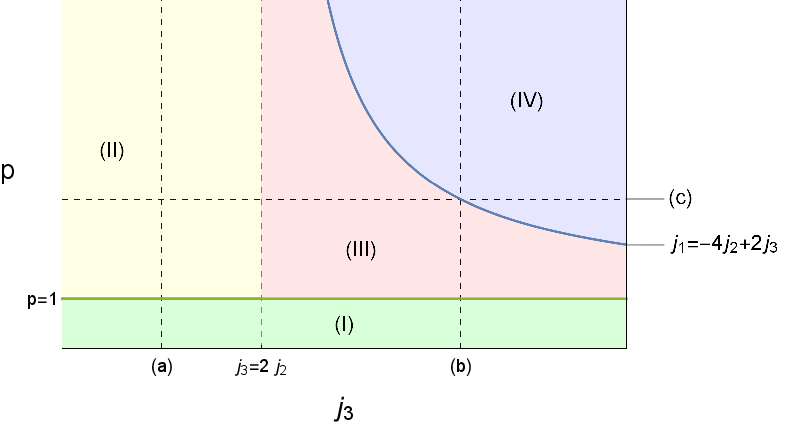}
	\caption{In this diagram we split the domain of the function $\mu_c(j)$ in convenient regions for $d=2$ and $j_2\neq 0$. Each colored patch corresponds to a region of the domain where the function is analytic. For an arbitrary fixed value of $j_2$, we have drawn the dashed line that splits the diagram between the regions $j_3<2j_2$ and $j_3>2j_2$. The lines $p=1$ and $j_1=-4j_2+2j_3$ are also explicitly shown. With the exception of the line $p=1$, by varying the $j$-parameters we can cross these regions maintaining $g\leq g_c$. These crossings configure further gapless-to-gapless phase transitions governed by the discontinuity of fist or second order derivatives of $\mu_c(j)$. The dashed lines (a), (b), and (c) are shown to mark the domain of the slices of the surface $g_c(j)$ that are depicted in Figures \ref{alldiagrams2d} \protect\subref{diagram2d-(a)}-\protect\subref{diagram2d-(c)}.}
	\label{diagram2d}
\end{figure}

In spite of the parameter space being four-dimensional, we can visualize the possible transitions among the regions listed above for $g\leq g_c$ by drawing a two-dimensional diagram $p$ {\it vs} $j_3$. In the diagram of Fig. \ref{diagram2d}, we have made explicit the regions $p<1$ and $p>1$. Also, without loss of generality, we can fix $j_2$ and draw the lines $j_3=2j_2$, and $j_1=2j_3-4j_2$. These lines are relevant due to the fact that crossing them by varying the $j$-parameters we transit among the regions listed in table (\ref{4.12}). In terms of $p$ and $j_3$, for $j_2$ fixed,  the curve $j_1=2j_3-4j_2$ is given by the curve $p=\frac{2j_3+4j_2}{2j_3-4j_2}$, which tends asymptotically to $p=1$ and $p=\infty$ as $j_3\rightarrow\infty$ and $j_3\rightarrow 2j_2$, respectively.

In each region, $\mu_c(j)$ is a smooth function of the $j$'s and no phase transition occurs by varying the parameters within the region. Then, in some regions, like $(II)$ for example, the ground state and the broken symmetry group are both rearranged when we vary $j$, but this rearrangement is completely continuous. Possible discontinuity can only happen by passing from one region to the others. 

\begin{figure}[h!]
	\subfloat[\label{diagram2d-(a)}]{ %
		\includegraphics[width=0.48\columnwidth]{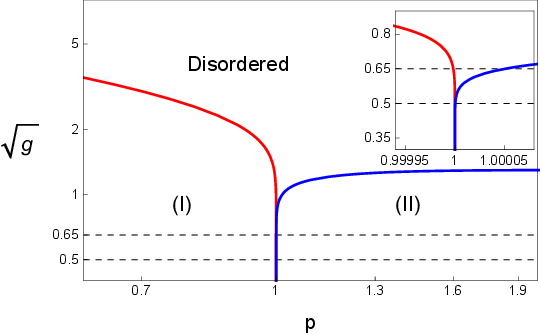} %
	}\hfill 
	\subfloat[\label{diagram2d-(b)}]{ %
		\includegraphics[width=0.48\columnwidth]{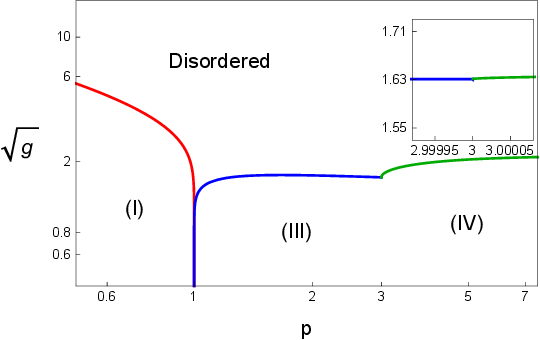} %
	}\hfill
	\subfloat[\label{diagram2d-(c)}]{ %
		\includegraphics[width=0.48\columnwidth]{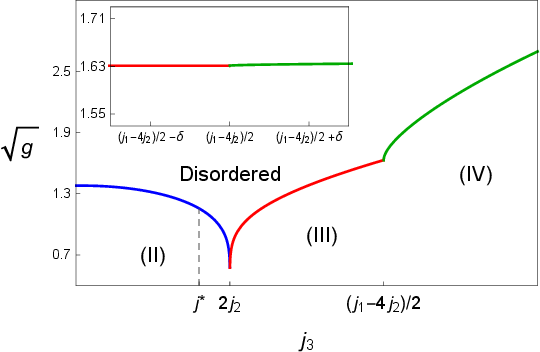} %
	}
	\caption{Figures \protect\subref{diagram2d-(a)}-\protect\subref{diagram2d-(c)} represent relevant slices of the critical surface $g_c(j)$ for $d=2$. To better localize these slices, we referred their domain to the dashed lines (a), (b), and (c) of Figure (\ref{diagram2d}). For the figure \protect\subref{diagram2d-(a)} we used the values $j_{2}=0.1$ and $j_{3}=0.11$; for the \protect\subref{diagram2d-(b)} we used $j_{2}=0.1$ and $j_{3}=0.4$, and for the \protect\subref{diagram2d-(c)} we used $j_{2} =  0.1$ and $p=3$. The inset in \protect\subref{diagram2d-(a)} shows that the apparently vertical line at $p=1$ corresponds in fact to two lines that touch only at $g_c=0$. The insets in \protect\subref{diagram2d-(b)}  and \protect\subref{diagram2d-(c)} show that both crossings at $(III\bigcap IV)$ have continuous first derivative. The separation $\delta$ from the point $j_3=\frac{j_1-4j_2}{2}$ is of the order $\delta \sim 10^{-5} j_{3}$.}
	\label{alldiagrams2d}
\end{figure}

From the diagram of Figs. \ref{alldiagrams2d} \subref{diagram2d-(a)} and \ref{alldiagrams2d} \subref{diagram2d-(b)}, we see that we cannot pass from region $(I)$ to the other regions maintaining $g<g_c$, i.e, without disordering the system. Therefore, we can discuss the passage from region $(II)$ to $(III)$, when $j_3$ and $j_2$ are varied and we go from $j_3<2j_2$ to $j_3>2j_2$. Also, we can discuss the passage from $(III)$ to $(IV)$ maintaining $j_3>2j_2$ by crossing the line $j_1=-4j_2+2j_3$. The critical chemical potential $\mu_c(j)$ is continuous through the intersections  $(II\bigcap III)$ and $(III\bigcap IV)$. Then, possible $n$-order phase transitions will occur if the $(n-1)$-th derivatives of $\mu_c$ fail to be continuous through the intersections.

It is straightforward to compute the discontinuities in the derivatives of $\mu_c$ through the intersections. We just show some of the possible discontinuities for the sake of clearness. For instance, in the transition ($II\leftrightarrow III$), we get
\begin{eqnarray}
D\left(\frac{\partial\mu_c}{\partial j_3}\right)\Big|_{\left(II\bigcap III\right)}&\equiv&\left(\frac{\partial\mu_c}{\partial j_3}\right)\Big|_{\left(II,j_3=2j_2\right)}-\left(\frac{\partial\mu_c}{\partial j_3}\right)\Big|_{\left(III,j_3=2j_2\right)}\\
&=&\left(\frac{j_1-8j_2}{4j_2}\right)^2.
\label{4.13}
\end{eqnarray}
The point $j_1\neq 8j_2$,  where the derivative would be continuous, is not in the region $p>1$ and cannot be attained through the transition. Thus, we conclude that $\left(\frac{\partial\mu_c}{\partial j_3}\right)$ is always discontinuous through the intersection $(II\bigcap III)$. 

The transition $(III\leftrightarrow IV)$ occurs through the line $j_1=-4j_2+2j_3$. This line is always in the region $j_3>2j_2$ and can be crossed varying any $j$ parameter. All first derivatives $\frac{\partial\mu_c}{\partial j_1}$ are continuous at $j_1=-4j_2+2j_3$, whereas for second derivatives we get 
\begin{eqnarray}
D\left(\frac{\partial^2\mu_c}{\partial j^2_1}\right)_{III\bigcap IV}=\frac{1}{2j_2},\label{4.16}
\end{eqnarray}
with similar second order discontinuities for the other derivatives.

To summarize our discussion, we conclude that besides the gapped-to-gapless phase transition occurring when we cross the surface $g_c(j)$ varying $g$ or $j$, we also encounter continuous gapless-to-gapless phase transitions between the ordered phases by varying the $j$-parameters below the surface $g_c(j)$.

Using the expression for the gap equation (\ref{4.1}) it is possible to generate numerically the three-dimensional surface $g_c(j_1,j_2,j_3)$ that separates the disordered region, $g>g_c$, from the ordered one, $g\leq q_c$. In Figs \ref{alldiagrams2d} \subref{diagram2d-(a)}, \ref{alldiagrams2d} \subref{diagram2d-(b)} and \ref{alldiagrams2d} \subref{diagram2d-(c)},  we show some relevant slices of this surface to better visualize how the regions of table (\ref{4.12}) are connected and the possible routes to phase transitions.

Between all the regions of the $(g~vs~j)$-diagram, the curve $g=g_c$ presents change of behavior. We can investigate the analytic properties of these curves and relate non-analytic points of the curves $g_c(j)$ with those of $\mu_c(j)$, which we know how to calculate analytically. To address this point, we recall our definition of  $\mathcal{O}\left({\bf q},j\right)\equiv \mu_c(j)-J({\bf q},j)$ and write $g^{-1/2}_c$ as
\begin{equation}
g^{-1/2}_c=\int\frac{d^d{\bf q}}{2\left(\mathcal{O}\left({\bf q},j\right)\right)^{1/2}}.\label{4.17}
\end{equation}
By proceeding as before, we conclude that the derivatives
\begin{equation}
\frac{dg_c}{dj}=g_c^{3/2}\int\frac{d^d{\bf q}}{2\left(\mathcal{O}\left({\bf q},j\right)\right)^{3/2}}\frac{\partial \mathcal{O}}{\partial j}\label{4.18}
\end{equation}
and
\begin{equation}
\frac{d^2g_c}{dj^2}=\frac{3}{2}g_c^{-1}\left(\frac{dg_c}{dj}\right)^{2}-\frac{3}{2}g_c^{3/2}\int\frac{d^d{\bf q}}{2\left(\mathcal{O}\left({\bf q},j\right)\right)^{5/2}}\left(\frac{\partial \mathcal{O}}{\partial j}\right)^2+g_c^{3/2}\int\frac{d^d{\bf q}}{2\left(\mathcal{O}\left({\bf q},j\right)\right)^{3/2}}\frac{\partial^2 \mathcal{O}}{\partial j^2}\label{4.19}
\end{equation}
are IR convergent. Now, since $\mathcal{O}\left({\bf q},j\right)\equiv \mu_c(j)-J({\bf q},j)$, with $J({\bf q},j)$ being linear in the $j$'s, the smoothness properties of $\mathcal{O}\left({\bf q},j\right)$ follow uniquely from $\mu_c(j)$. In its turn, $\mu_c(j)$ is smooth inside each region of the table (\ref{4.12}), with possible discontinuities in first or second derivatives at the intersection of these regions. We then see that $\mathcal{O}({\bf q},j)$ is continuous, but its derivatives with respect to $j$ can be discontinuous at the intersections, i.e., the discontinuities of the derivatives of $\mathcal{O}({\bf q},j)$ are directly related to those of $\mu_c(j)$. This line of reasoning gives an one-to-one connection between the non-analytic points in the curves $g_c(j)$ with those of $\mu_c(j)$.

From this analysis,  we can explain the smoothness of the curves $g_c(j)$ inside each region and the cusps at the intersections. For instance, from (\ref{4.13}) and (\ref{4.16}), we conclude that $\frac{dg_c}{dj_3}$ is discontinuous at $(II\bigcap III)$ and $\frac{d^2g_c}{dj^2_1}$ is discontinuous at $(III\bigcap IV)$. This behavior was studied numerically, as shown in Fig. \ref{alldiagrams2d} \subref{diagram2d-(c)}.

\vspace{0.3cm}
\underline{$d=3$ with $j_{2} \neq 0$}
\vspace{0.3cm}

For $d=3$ we use the information of the tables (\ref{0.37}) and (\ref{0.38}) to calculate $\mu_c(j)=J({\bf q}^c)$ in all relevant regions:
\begin{equation}
\mu_c(j)=
\begin{cases}
\; 6(j_1-j_2-2j_3) &\!  \text{for}~~~p<1~~~\text{or for}~~~p=1~\text{and}~j_2\neq 0~~~(I) \\
\; 6j_2 + \frac{3j_1^2}{4\left(j_2 + j_3\right)} &\!  \text{for}~~~p\geq 1~~~\text{and}~~~j_3\leq 2j_2~~~(II)\\
\; \frac{j_1^2+24j^2_2}{4j_2} &\!  \text{for}~~~p\geq 1~~~\text{and}~~~j_3=2j_2~~~(II\bigcap III)\\
\; \frac{8j_1\left(2j_2-j_3\right)-8\left(j^2_2+2j_2j_3-2j^2_3\right)+j^2_1}{4j_2} &\!  \text{for}~~~p\geq 1,~~~j_3\geq 2j_2,~~\text{and}~~~j_1\geq 4j_3-4j_2~~~(III)\\
\; 2\left(j_1-3j_2+2j_3\right) &\!  \text{for}~~~p>1,~~~j_3\geq 2j_2,~~\text{and}~~~4j_3-4j_2\geq j_1\geq 4j_2~~~(IV)\\
\; \frac{j^2_1}{4j_2}-2j_2+4j_3 &\!  \text{for}~~~p>1,~~~j_3\geq 2j_2,~~\text{and}~~~j_1\leq 4j_2~~~(V).
\end{cases}
\label{4.21}
\end{equation}

\begin{figure}[h!]
	\includegraphics[scale=1]{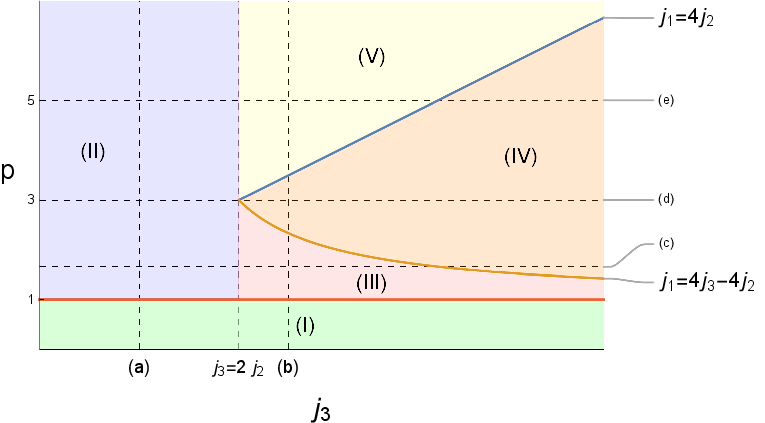}
	\caption{This is the analogue of the Fig. \ref{diagram2d} for $d=3$, where we split the domain of the function $\mu_c(j)$ in convenient regions. Again, for $j_2\neq 0$, each colored patch corresponds to a region of the domain where the function is analytic. For an arbitrary fixed value of $j_2$, we have drawn the dashed line that splits the diagram between the regions $j_3<2j_2$ and $j_3>2j_2$. The lines $p=1$, $j_1=4j_3-4j_2$, and $j_1=4j_2$ are also explicitly shown. By varying the $j$-parameters we can cross these regions maintaining $g\leq g_c$. Differently from the $d=2$ case, here we have $g_c\neq 0$ at $p=1$ and we can have a transition across $p=1$ maintaining $g\leq g_c$. These crossings configure further gapless-to-gapless phase transitions governed by the discontinuity of first or second order derivatives of $\mu_c(j)$. The dashed lines (a)-(e) are shown to mark the domain of the slices of the surface $g_c(j)$ that are depicted in Figs. \ref{alldiagrams3d} \protect\subref{diagram3d-(a)}-\protect\subref{diagram3d-(e)}.}
	\label{diagram3d}
\end{figure}

Again, we can visualize the possible transitions among the regions listed above for $g\leq g_c$ by drawing a two-dimensional diagram $p$ {\it vs} $j_3$ (Fig. \ref{diagram3d}). Then, we can have  the transitions $(I\leftrightarrow II)$, $(I\leftrightarrow III)$, $(II\leftrightarrow III)$, $(II\leftrightarrow IV)$, $(II\leftrightarrow V)$, $(III\leftrightarrow IV)$, and $(IV\leftrightarrow V)$. Analogously to what we have done for $d=2$, the nature of these transitions are characterized by the least-order derivative of $\mu_c$ that is discontinuous through the intersections. From  table (\ref{4.21}), it is straightforward to show that the first derivatives with respect to $j_2$ and $j_3$ are discontinuous in all transitions $(II\leftrightarrow any)$, whereas for the other transitions the discontinuities are in the second derivatives.

\begin{figure}[h!]
	\subfloat[\label{diagram3d-(a)}]{ %
		\includegraphics[width=0.48\columnwidth]{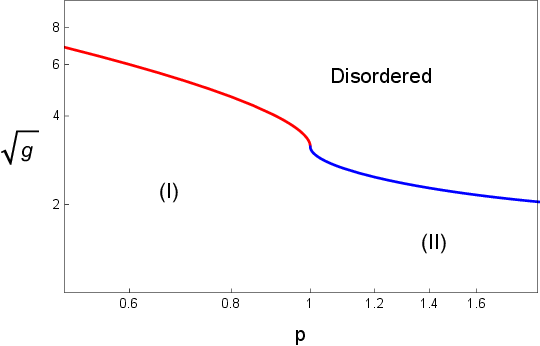}}\hfill 
	\subfloat[\label{diagram3d-(b)}]{ %
		\includegraphics[width=0.48\columnwidth]{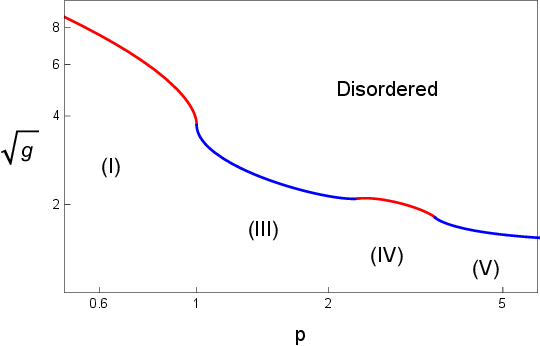}}\hfill
	\subfloat[\label{diagram3d-(c)}]{ %
		\includegraphics[width=0.48\columnwidth]{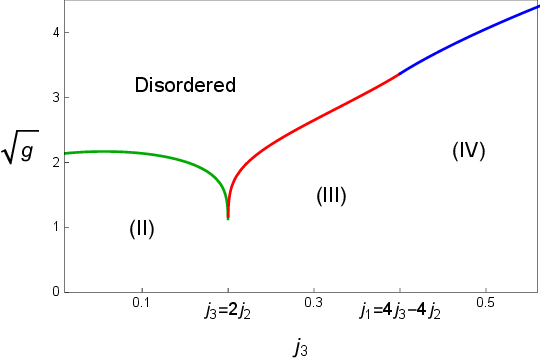}}\hfill
	\subfloat[\label{diagram3d-(d)}]{ %
		\includegraphics[width=0.48\columnwidth]{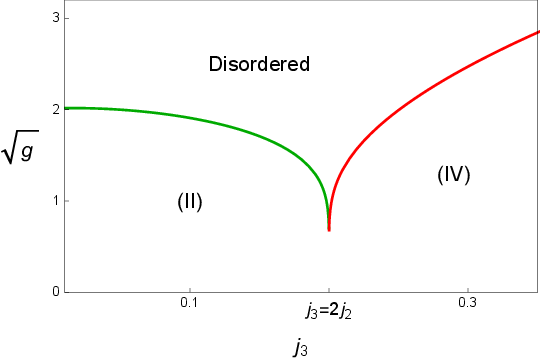}}\hfill
	\subfloat[\label{diagram3d-(e)}]{ %
		\includegraphics[width=0.48\columnwidth]{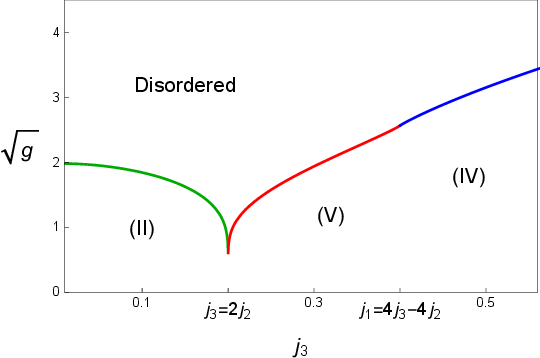}}
	\caption{Figures \protect\subref{diagram3d-(a)}-\protect\subref{diagram3d-(e)} represent relevant slices of the critical surface $g_c(j)$ for $d=3$. To better localize these slices, we referred their domain to the dashed lines (a)-(e) in the Figure (\ref{diagram3d}). All these diagrams were plotted using $j_{2} = 0.1$. For the  figures \protect\subref{diagram3d-(a)}  and \protect\subref{diagram3d-(b)} we used $j_{3}=0.11$ and $j_{3}=0.25$, respectively. For the figures \protect\subref{diagram3d-(c)}, \protect\subref{diagram3d-(d)}, and \protect\subref{diagram3d-(e)} we used $p=5/3$, $p=3$ and $p=5$, respectively.}
	\label{alldiagrams3d}
\end{figure}

To sum, the situation is analogous to $d=2$. We find a gapped-to-gapless transition in traversing the surface $g_c(j)$ and several gapless-to-gapless phase transitions in going through the intersections of the ordered phases ($g<g_c(j)$) in the regions depicted in table (\ref{4.21}). In Fig. \ref{alldiagrams3d}, we show some relevant slices of the surface $g_c(j_1,j_2,j_3)$. We can also verify through the analysis of the equations (\ref{4.18}) and (\ref{4.19}) that their analytic properties are in one-to-one correspondence with those of $\mu_c(j)$.

\vspace{0.3cm}
\underline{$d=3$ with $j_{2} = 0$}
\vspace{0.3cm}

\begin{figure}[h!]
		\subfloat[\label{diagramj2}]{ %
		\includegraphics[width=0.475\columnwidth]{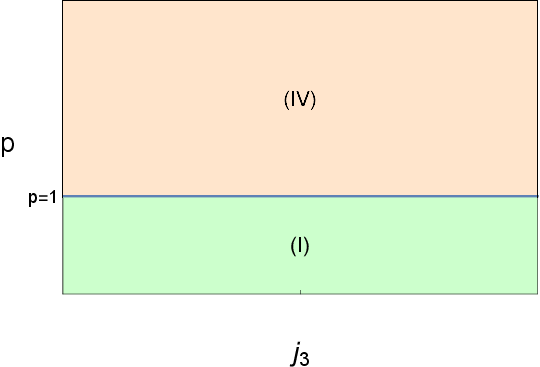}}\hfill 
		\subfloat[\label{diagramj2-(a)}]{ %
		\includegraphics[width=0.485\columnwidth]{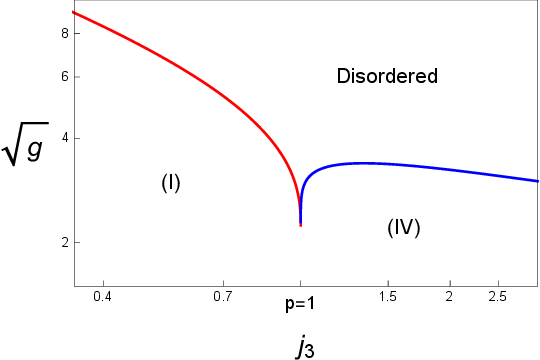}}\hfill
	\caption{\protect\subref{diagramj2} represents the regions defined by \eqref{89}. For the phase diagram in \protect\subref{diagramj2-(a)} we used $j_{3} = 0.25$.}
	\label{diagramj2both}
\end{figure}

The case with $j_2=0$ needs to be handled separately, since many of the regions in the diagram in Fig. \ref{diagram3d} are suppressed. In this case, the table \eqref{4.21} reduces to
\begin{equation}
	\mu_c(j)=
	\begin{cases}
		\; 6(j_1-2j_3) &\!  \text{for}~~~p<1~~~(I) \\
		\; 12j_3 &\!  \text{for}~~~p=1~~~(I \bigcap IV) \\
		\; 2\left(j_1+2j_3\right) &\!  \text{for}~~~p>1~~~(IV)
	\end{cases}.
	\label{89}
\end{equation}
This can also be seen by analyzing the diagram of Fig. \ref{diagram3d} with $j_{2} \rightarrow 0$. Notice that the curve $j_3=2j_{2}$ tends to $j_3=0$ and the region $(II)$ disappears. The curve $j_1=4j_{2}$, which in the plane $p-j_3$ is given by $p=\frac{j_3}{j_2}$, gets mapped to the point $p=\infty$ when $j_2\rightarrow 0$. Also, the curve $j_1=4j_3-4j_2$, which is given by $p=\frac{j_3}{j_3-j_2}$, merges with the line $p=1$. The regions defined by \eqref{89} are depicted in Fig. \ref{diagramj2both} \subref{diagramj2}. Contrary to the two dimensional case, for $d=3$ the system is able to order when $p=1$ with $j_{2} = 0$, since it presents a nonvanishing $g_{c}$, as can be seen more directly from the diagram in Fig. \ref{diagramj2both} \subref{diagramj2-(a)}. The gapless-to-gapless phase transition as we go from the region $(I)$ to $(IV)$ by crossing the line $p=1$ (with $g<g_{c}$), exhibits first order discontinuities in the derivatives $\frac{\partial \mu}{\partial j_{1}}$ and $\frac{\partial \mu}{\partial j_{3}}$. The transition region $p=1$ is special in that the system presents the exotic behavior discussed in the end of Sec. \ref{sec4b}.


\section{SSB and the nature of the Nambu-Goldstone bosons\label{sec6}}

In this section we investigate in more detail the phases that appear in the model (\ref{4.24}) by elucidating the symmetries involved in the mechanism of phase transitions, as well as the way these symmetries can break and generate Nambu-Goldstone excitations. For this purpose we revisit the cases we have analyzed  in the previous section.

To begin the discussion we rewrite the total Hamiltonian (\ref{1.5}) as
\begin{equation}
\mathcal{H}=\frac{1}{2}\sum_{{\bf q}\,\in\,\left[{\bf q}^c\right]}\left(g\Pi_{\bf{q}}\Pi_{-\bf{q}}+\frac{1}{g}~\omega_{\bf q}^2 S_{\bf{q}} S_{-\bf{q}}\right)~+\sum_{\bf q\,\notin\,\left[{\bf q}^c\right]} \omega_{\bf q}\left(a_{\bf q}^{\dagger}a_{\bf q}+\frac{1}{2}\right),\label{6.1}
\end{equation}
where we have used the definitions of creation and annihilation operators (\ref{1.9}) and (\ref{1.10}) only for the non-critical modes. 
This Hamiltonian describes either the disordered phase, $\mu>\mu_c$, or the ordered and modulated ones, for $\mu=\mu_c$. To analyze the mechanism of phase transitions, we focus in the first term that we call the critical mode Hamiltonian:
\begin{equation}
\mathcal{H}_{\left[{\bf q}^c\right]}\equiv\frac{1}{2}\sum_{{\bf q}\,\in\,\left[{\bf q}^c\right]}\left(g\Pi_{\bf{q}}\Pi_{-\bf{q}}+\frac{1}{g}\omega_{\bf q}^2 S_{\bf{q}} S_{-\bf{q}}\right).\label{6.2}
\end{equation}
This Hamiltonian corresponds to a system of decoupled harmonic oscillators with identical frequencies $\omega_{{\bf q}}$ for ${\bf q}\in \left[{\bf q}^c\right]$. Therefore, the Hamiltonian is invariant under permutations of the harmonic oscillator dynamical variables. In the following we discuss the general properties of the ordered phases for each possible set of $\left[{\bf q}^c\right]$.
 
 
\subsection{$p\leq 1$ and $j_2\neq 0$}

In this region the critical momentum is given by ${\bf q}^c=\left(0,\ldots,0\right)$. To investigate the properties of the phase transitions we turn on a constant external magnetic field, which only couples to the critical mode, by adding the term $h\sum S_{\bf r}$ to the Hamiltonian. Therefore, we simply get
\begin{equation}
\mathcal{H}_{0}\equiv\frac{g}{2}(\bar{P})^2+\frac{1}{2g}\omega_{0}^2 (\bar{X})^2-\sqrt{N}h\bar{X},
\label{3.1a}
\end{equation}
with
\begin{equation}
\bar{P}\equiv \Pi_{0}~~~\text{and}~~~\bar{X}\equiv S_{0}=\frac{1}{\sqrt{N}}\sum_{\bf r} S_{\bf r}.
\label{3.2}
\end{equation}

Completing to a square, the Hamiltonian can be written as
\begin{eqnarray}
\mathcal{H}_{0}&=&\frac{g}{2}(\bar{P})^2+\frac{1}{2g}\omega_{0}^2 \left(\bar{X}-\frac{\sqrt{N}h}{\mu-J(0)}\right)^2-\frac{N}{2}\frac{h^2}{(\mu-J(0))}.
\label{3.6}
\end{eqnarray}
Let us investigate some properties of the eigenvalues of $\bar{X}$. For this purpose, as we have done in (\ref{2.11}), we consider the spherical constraint in the Fourier space and separate the zero critical mode from the other modes. The mean value is calculated with respect to eigenstates of $\bar{X}$ and the Fock vacuum for the other modes, i.e., $a_{{\bf q}}\left|m\rangle\right.=0$ and $\bar{X}\left|m\rangle\right.=\sqrt{N}m\left|m\rangle\right.$. Denoting by $\left|m\rangle_{\bf r}\right.$ the eigenstate of $S_{\bf r}$ with eigenvalue $m$, we have explicitly $\left|m\rangle\right.=\bigotimes_{\bf r}\left|m\rangle_{\bf r}\right.$. Then, we obtain
\begin{eqnarray}
\left.\langle m\right|(\bar{X})^2+\sum_{{\bf q}\neq 0}S_{\bf q}S^\dagger_{\bf q}\left|m\rangle\right.&=&N.\label{3.7}
\end{eqnarray}
Using the relations (\ref{1.9}) and (\ref{1.10}), we can rewrite equation (\ref{3.7}) in terms of the creation and annihilation operators:
\begin{eqnarray}
N m^2+\sum_{{\bf q}\neq 0}\left(\frac{g}{2\omega_q}\right)\left.\langle m\right|\left(a_{{\bf q}}+a^\dagger_{{\bf q}}\right)^2\left|m\rangle\right.&=&N.
\label{3.10}
\end{eqnarray}
Using the algebra of ladder operators and the fact that the lowering operators annihilate the states, we obtain a sum $\sum_{{\bf q}\neq 0}\left(\frac{g}{2\omega_q}\right)$ in the second term on the left-hand side. This sum gives the mean occupation number of the non-critical modes in the states $\left|m\rangle\right.$. Denoting the fraction of non-critical modes by $m_f$, we have 
\begin{eqnarray}
N\left(m^2+m_f^2\right)&=&N,
\label{3.11}
\end{eqnarray}
which implies that $-1\leq m\leq 1$. Therefore, the eigenvalues of $\bar{X}$ are limited to the interval $\left[-\sqrt{N},\sqrt{N}\right]$, and the spectrum of the conjugate momentum $P$ is quantized as
\begin{equation}
p=\frac{n\pi}{\sqrt{N}},~~~\text{with}~~~n\in \mathbb{Z}.\label{3.12} 
\end{equation}
For convenience, we define rescaled operators $P\equiv\sqrt{N}\bar{P}$ and $X\equiv\frac{\bar{X}}{\sqrt{N}}$, whose spectrum does not depend on $N$. In terms of these intensive operators, the Hamiltonian is written as
\begin{eqnarray}
\mathcal{H}_{0}&=&\frac{g}{2N}(P)^2+\frac{N}{2g}~\omega_{0}^2 \left(X-\frac{h}{\mu-J(0)}\right)^2-\frac{N}{2}\frac{h^2}{(\mu-J(0)) },
\label{3.13}
\end{eqnarray}
which describes an harmonic oscillator with frequency $\omega_0$ with the coordinate $X$ oscillating around the position $\frac{h}{\mu-J(0)}$. So, the ground state wave function is a Gaussian given by
\begin{equation}
\Psi[m]\sim  \exp\left[- \frac{N \omega_{0}}{2g}\left(m-\frac{h}{\mu-J(0)}\right)^2\right].
\label{3.14}
\end{equation}
The Gaussian is characterized by the dispersion 
\begin{equation}
\sigma^2\equiv \frac{g}{N\omega_{0}}
\label{3.15}
\end{equation}
and is centered about 
\begin{equation}
\tilde{m}\equiv\frac{h}{\mu-J(0)}.
\label{3.16}
\end{equation}
This is the expected value of the magnetization, which can be calculated from the partition function as $m_N=\lim_{\beta\rightarrow\infty}-\frac{1}{\beta}\frac{\partial}{\partial h}\ln Z$.
Then, if we take a specific sequence of limits, we obtain
\begin{equation}
\lim_{\overset{h\rightarrow 0}{\mu\rightarrow\mu_c}}\lim_{N\rightarrow\infty}\left|\Psi\left[m\right]\right|^2~~\rightarrow~~ \delta(m-\tilde{m}).\label{3.17}
\end{equation}
On the other hand, if we interchange the order of the limits, we get
\begin{equation}
\lim_{N\rightarrow\infty} \lim_{\overset{h\rightarrow 0}{\mu\rightarrow\mu_c}}\left|\Psi\left[m\right]\right|^2=\text{constant}.\label{3.18}
\end{equation}

The singularity due to the non-commutation of the limits above is a signal of spontaneous symmetry breaking. In fact, turning off the external field $h$, we see that for $\omega_{0}=0$ the charge $P$ is conserved. From (\ref{1.16}), for finite $N$, the energy spectrum in the critical-mode sector is proportional to $\frac{n^2}{N}$, and we have a unique ground state, for which $n=0$. When $N\rightarrow\infty$ all the excited critical modes states, $n\neq 0$, become degenerate with the ground state $n=0$, ensuing the spontaneous breaking of the $U(1)$ symmetry generated by $P$. 

To further investigate this scenario, let us consider the quasi-degenerate excited states for finite $N$. Since the eigenvalues of $\bar{X}$ are confined between $-\sqrt{N}$ and $\sqrt{N}$ and we are interested in the large $N$ limit, we can allow these eigenvalues to be periodic variables with period $2\sqrt{N}$, which in turn implies that the eigenvalues of $X$ can be taken periodic with period 2. Then, the $P$ eigenstates
\begin{equation}
\Psi_{p}[m]=e^{imp}\label{3.19}
\end{equation}
are also energy eigenstates. Since the eigenvalues of $X$ are periodic, $m\sim m+2$, the $p$ eigenvalues are quantized as $p=n\pi$, with $n\in\mathbb{Z}$. The symmetry is implemented by the unitary operator $U(m)=e^{imP}$, and all the energy eigenstates are symmetric in the sense that $U(m)\left|p\rangle\right.=e^{imp}\left|p\rangle\right.$. In the limit $N\rightarrow\infty$ all the states (\ref{3.19}) are degenerate with the vacuum $n=0$. However, since these states are non-local in the $m$ variables, they become unstable under arbitrarily small perturbations of external field $h$. Forming linear combinations of the $P$ eigenstates it is possible to build localized states as
\begin{equation}
\left|m\rangle\right.=\sum_{p} e^{i m p}\left|p\rangle\right..\label{3.20}
\end{equation}
For finite $N$ these are not energy eigenstates, but since the energy gap goes as $1/N$, they become degenerate with the ground state in the large $N$ limit, and they are stable under small external field perturbations. Therefore, the eigenstates of $X$ form a degenerate set of stable ground states in the ordered phase. The elements of this set are connected through the action of the spontaneously broken symmetry operator $U(m)$:
\begin{equation}
U(m^{\prime})\left|m\rangle\right.=\left|m+m^{\prime}\rangle\right.. \label{3.21}
\end{equation}

It is worth emphasizing that in the thermodynamic limit there is no phase where the symmetry above is exactly realized: in the disordered phase, $\mu\neq\mu_c$, the symmetry is explicitly broken by the presence of the quadratic potential proportional to $X^2$, and in the ordered phase, $\mu=\mu_c$, the symmetry is spontaneously broken. This can also be seen when we consider the two different order of limits $\lim_{h\rightarrow 0}\lim_{N\rightarrow\infty}$ and $\lim_{N\rightarrow\infty}\lim_{h\rightarrow 0}$ in the ground state wave function (\ref{3.14}) without taking $\mu\rightarrow\mu_c$ simultaneously with $h\rightarrow 0$. We can easily verify that these two limits commute and and we do not have spontaneous ordering. In fact, we will be always in the phase with explicit broken symmetry with the unique ground state being a sharply peaked Gaussian centered at $m=0$, as enforce by the potential $X^2$.


\subsection{$p>1$ and $j_3\neq 2j_2$}

From the list of possible regions in parameter space (\ref{0.37}) and (\ref{0.38}) it is possible to see that in any dimension, except for the cases $p=1$ with $ j_2=0$ and $p>1$ with  $j_3=2j_2$, given a point in the parameter space, the critical momentum is unique up to permutations and sign change operations of its components. To describe the mechanism of phase transition in all these regions we parallel the discussion of the previous case with ${\bf q}^c=0$.

Now we add a modulated external field to the critical mode Hamiltonian (\ref{6.2}) of the form
\begin{equation}
h_{\bf r}=\sum_{{\bf q}\,\in\,\left[{\bf q}^c\right]}\left(h^{(1)}_{{\bf q}}\cos\left({\bf q}\cdot{\bf r}\right)+h^{(2)}_{{\bf q}}\sin\left({\bf q}\cdot{\bf r}\right)\right), \label{3.1}
\end{equation}
giving the following contribution to the Hamiltonian
\begin{equation}
-\sum_{\bf r} h_{\bf r}S_{\bf r}=-\sqrt{N}\sum_{{\bf q}\,\in\,\left[{\bf q}^c\right]}\left(h^{(1)}_{{\bf q}}\,\text{Re}S_{{\bf q}}+h^{(2)}_{{\bf q}}\,\text{Im}S_{{\bf q}}\right).
\label{3.22}
\end{equation}
As before, the external fields couples only to the critical modes, and then, the Hamiltonian we consider now is 
\begin{eqnarray}
\mathcal{H}_{\left[{\bf q}^c\right]}&=&\sum_{{\bf q}\,\in\,\left[{\bf q}^c\right]}\sum_{i=1}^2\left(\frac{g}{2N}\left(P^{(i)}_{{\bf q}}\right)^2+\frac{N}{2g}~\omega_{{\bf q}}^2 \left(X^{(i)}_{{\bf q}}-\frac{h^{(i)}_{{\bf q}}}{\mu-J\left({\bf q}\right)}\right)^2 - \frac{N}{2}\frac{\left(h^{(i)}_{{\bf q}}\right)^2}{\left(\mu-J\left({\bf q}\right)\right) }\right),
\label{3.31}
\end{eqnarray}
where we have defined again intensive operators
\begin{equation}
P^{(1)}_{{\bf q}}\equiv \sqrt{N}\,\text{Re}\Pi_{{\bf q}}~~~\text{and}~~~P^{(2)}_{{\bf q}}\equiv \sqrt{N}\,\text{Im}\Pi_{{\bf q}},
\label{3.24}
\end{equation}
and also
\begin{equation}
X^{(1)}_{{\bf q}}\equiv \frac{1}{\sqrt{N}}\,\text{Re}S_{{\bf q}}~~~\text{and}~~~X^{(2)}_{{\bf q}}\equiv \frac{1}{\sqrt{N}}\,\text{Im}S_{{\bf q}}.
\label{3.25}
\end{equation}
Since, $S^\dagger_{{\bf q}}=S_{-{\bf q}}$ and $\Pi^\dagger_{{\bf q}}=\Pi_{-{\bf q}}$, we get 
\begin{eqnarray}
P^{(i)}_{-{\bf q}}&=&\left(-1\right)^{i+1}P^{(i)}_{{\bf q}},\\\label{6.26}
X^{(i)}_{-{\bf q}}&=&\left(-1\right)^{i+1}X^{(i)}_{{\bf q}}.\label{6.27}  
\end{eqnarray}

Due to these constraints, we have the Hamiltonian for $\mathsf{N}_c$ decoupled harmonic oscillator instead of only one as in (\ref{3.13}). It is worth to notice that the external fields break the permutation symmetry of the Hamiltonian and introduce a distinction between the oscillators. The ground state wave function is then the product $\Psi[m^{(1)},m^{(2)}]=\prod_{\overset{{{\bf q}}\,\in\,\left[{\bf q}^c\right]/\mathbb{Z}_2}{i=1,2}}\Psi[m^{(i)}_{{\bf q}}]$ of $\mathsf{N}_c$ independent Gaussian functions
\begin{equation}
\Psi[m^{(i)}_{{\bf q}}]\sim  \exp\left[- \frac{N \omega_{{\bf q}}}{g}\left(m^{(i)}_{{\bf q}}-\frac{h^{(i)}_{{\bf q}}}{\mu-J\left({\bf q}\right)}\right)^2\right],
\label{3.32}
\end{equation}
with the dispersion 
\begin{equation}
\sigma^2\equiv \frac{g}{2N\omega_{{\bf q}}}
\label{3.33}
\end{equation}
and are centered about 
\begin{equation}
\tilde{m}^{(i)}_{{\bf q}}\equiv \frac{h^{(i)}_{{\bf q}}}{\mu-J\left({\bf q}\right)}.
\label{3.34}
\end{equation}
Again, we have the non-commutativity of the limits
\begin{eqnarray}
\lim_{\overset{h\rightarrow 0}{\mu\rightarrow\mu_c}}\lim_{N\rightarrow\infty}\left|\Psi\left[m\right]\right|^2~~&\rightarrow&~~ \delta(m-\tilde{m})\\
\lim_{N\rightarrow\infty} \lim_{\overset{h\rightarrow 0}{\mu\rightarrow\mu_c}}\left|\Psi\left[m\right]\right|^2&=&\text{constant},\label{3.35}
\end{eqnarray}
which signalizes the spontaneous breaking of the group $U(1)^{\mathsf{N}_c}$, generated by the conserved charges $P^{(i)}_{{\bf q}}$. 

The ordered ground state of the system is of the form
\begin{equation}
|\Omega\rangle_{\left[{\bf q}^c\right],m}=|m^{(1)},m^{(2)}\rangle_{\left[{\bf q}^c\right]}\bigotimes_{{\bf q}\,\notin\,\left[{\bf q}^c\right]}|0\rangle,\label{6.50}
\end{equation}
where $|m^{(1)},m^{(2)}\rangle_{\left[{\bf q}^c\right]}$ is the ground state of $\mathcal{H}_{\left[{\bf q}^c\right]}$ for $\mu=\mu_c$ and $|0\rangle$ is the Fock ground state for the non-critical modes, i.e., $a_{{\bf q}}\left|0\rangle\right.=0$. As discussed above, $|m^{(1)},m^{(2)}\rangle_{\left[{\bf q}^c\right]}$ can be constructed as a tensor product of eigenstates of $X^{(i)}_{{\bf q}}$, with ${\bf q}\in [{\bf q}^c]$. Defining $|m^{(i)}_{{\bf q}}\rangle$ such that $X^{(i)}_{{\bf q}}|m^{(i)}_{{\bf q}}\rangle=m^{(i)}_{{\bf q}}|m^{(i)}_{{\bf q}}\rangle$, we have from (\ref{6.27}) that $m^{(i)}_{-{\bf q}}=\left(-1\right)^{i+1}m^{(i)}_{{\bf q}}$. Then, we can write
\begin{equation}
|m^{(1)},m^{(2)}\rangle_{\left[{\bf q}^c\right]}=\bigotimes_{{\bf q}\,\in\,\left[{\bf q}^c\right]/\mathbb{Z}_2}|m^{(1)}_{{\bf q}}\rangle\otimes|m^{(2)}_{{\bf q}}\rangle.
\label{6.51}
\end{equation}
In turn, it may be convenient to express $|m^{(i)}_{{\bf q}}\rangle$ in terms of the eigenstates $|m\rangle_{{\bf r}}$ of $S_{{\bf r}}$. In this case, we show explicitly that
\begin{eqnarray}
|m^{(1)}_{{\bf q}}\rangle&=&\bigotimes_{{\bf r}}|2m^{(1)}_{{\bf q}}\cos\left({\bf q}\cdot{\bf r}\right)\rangle_{{\bf r}},\\\label{6.52}
|m^{(2)}_{{\bf q}}\rangle&=&\bigotimes_{{\bf r}}|2m^{(2)}_{{\bf q}}\sin\left({\bf q}\cdot{\bf r}\right)\rangle_{{\bf r}}.\label{6.53}
\end{eqnarray}
For instance, from (\ref{3.25}) we have
\begin{eqnarray}
X^{(1)}_{{\bf q}}|m^{(1)}_{{\bf q}}\rangle&=&\frac{1}{N}\sum_{\bf r}S_{\bf r}\cos\left({\bf q}\cdot{\bf r}\right)\bigotimes_{{\bf r}^\prime}|2m^{(1)}_{{\bf q}}\cos\left({{\bf q}}\cdot{\bf r}^\prime\right)\rangle_{{\bf r}^\prime}\nonumber\\
&=&\frac{2}{N}\sum_{{\bf r}}m^{(1)}_{{\bf q}}\cos^2\left({\bf q}\cdot{\bf r}\right)|m^{(1)}_{{\bf q}}\rangle\nonumber\\
&=&m^{(1)}_{{\bf q}}|m^{(1)}_{{\bf q}}\rangle.\label{6.54}
\end{eqnarray}

By analyzing the spherical constraint in momentum space, we again find the range of the eigenvalues of the operators $X^{(i)}_{{\bf q}}$. The mean value is calculated with respect to the ordered ground state $\left|\Omega\rangle_{\left[{\bf q}^c\right],m}\right.$. Using the definitions (\ref{6.50})-(\ref{6.53}), and repeating the steps that led us to equation (\ref{3.11}), we obtain 
\begin{eqnarray}
N\left(\sum_{{\bf q}\,\in\,\left[{\bf q}^c\right]}\sum_{i=1}^2\left(m^{(i)}_{{\bf q}}\right)^2+m_f^2\right)&=&N,
\label{3.29}
\end{eqnarray}
with $m_f$ being the fraction of non-critical modes. This equation implies that the eigenvalues of $X^{(i)}_{{\bf q}}$ are  again limited to the interval $-1\leq m^{(i)}_{{\bf q}}\leq 1$ and the spectrum of the conjugate momenta $P^{(i)}_{{\bf q}}$ are quantized as
\begin{equation}
p^{(i)}_{{\bf q}}=n^{(i)}_{{\bf q}}\pi,~~~\text{with}~~~n^{(i)}_{-{\bf q}}=(-1)^{i+1}n^{(i)}_{{\bf q}},~~n^{(i)}_{{\bf q}}\in \mathbb{Z}.\label{3.30} 
\end{equation}

Given a set $\left[{\bf q}^c\right]$, there is an infinite number of ordered ground states labeled by $m^{(i)}$, which in fact corresponds to a set of labels $m^{(i)}_{{\bf q}}$ with ${\bf q}\in\left[{\bf q}^c\right]$. The $P^{(i)}_{{\bf q}}$ eigenstates $|p^{(1)}_{{\bf q}},p^{(2)}_{{\bf q}}\rangle$  are symmetric states, but they are unstable under small perturbations of the local external fields (\ref{3.1}). The critical mode part of the ordered ground state can also be constructed from the linear combination of the symmetric eigenstates:
\begin{equation}
|m^{(1)},m^{(2)}\rangle_{\left[{\bf q}^c\right]}
=\sum_{\{p^{(i)}_{{\bf q}}\}}\exp\left(i\sum_{{\bf q}\,\in\,\left[{\bf q}^{c}\right]/\mathbb{Z}_2}\sum_{i=1}^2m^{(i)}_{{\bf q}}p^{(i)}_{{\bf q}}\right)|p^{(1)}_{{\bf q}},p^{(2)}_{{\bf q}}\rangle,\label{3.36}
\end{equation}
which become degenerate with the ground state in the large $N$ limit and are stable under small external field perturbations. 
These ground states are connected by the broken symmetry operator 
\begin{equation}
U(m^{(1)},m^{(2)})=\exp\left(i\sum_{{\bf q}\,\in\,\left[{\bf q}^{c}\right]/\mathbb{Z}_2}\sum_{i=1}^2 m^{(i)}_{{\bf q}}P^{(i)}_{{\bf q}}\right).
\label{3.24a}
\end{equation}
Indeed, we can immediately check that
\begin{equation}
U(m^{\prime (1)},m^{\prime (2)})\left|m^{(1)},m^{(2)}\rangle_{\left[{\bf q}^c\right]}\right.=\left|m^{(1)}+m^{\prime (1)},m^{(2)}+m^{\prime (2)}\rangle_{\left[{\bf q}^c\right]}\right.. \label{3.37}
\end{equation}
As discussed in the end of the last section, there is no phase where the symmetry above is exactly realized.


\subsection{Exotic Cases: $p=1$ with $j_2=0$ and $p>1$ with $j_3=2 j_2$}

For the cases previously considered, the critical modes are specified by a set $[{\bf q}^c]$, which is generated by considering permutations and sign changes of a given ${\bf q}^c$. However, in the regions considered in this section, there is a huge degeneracy in ${\bf q}^c$ that cannot be reached just from permutations and changes of sign of the components. Moreover, the number of conserved charges goes to infinity in the thermodynamic limit.

Most of the expressions of the previous section, like the Hamiltonian (\ref{3.31}) and the ground states (\ref{6.51}), are formally the same for this case, except that the sum over the critical modes ranges over all possible sets $\left[{\bf q}^c\right]$. For example, in the two-dimensional case with $p=1$ and $j_2=0$, each specific critical mode $(0,q)$ gives rise to a set $[(0,q)]$ (see table \ref{0.28}). By varying $q$ one produces a large number of sets. Due to this extra degeneracy in the critical momenta we can go further and define quasi-local quantities, which provide a clearer view of the situation. 

On the surface $p=1$ with $j_2=0$ in $d$ dimensions, for instance, we have ${\bf q}^c=\left(0,\ldots,q_i,0,\ldots,0\right)$ with $i=1,\ldots,d$ and $q_i$ belonging to the first Brillouin zone. Then, we can define the operators
\begin{eqnarray}
X^{(1)}_i(x^i)&\equiv&\sum_{{q}^i}X^{(1)}_{\left(0,\ldots,q^i,0,\ldots,0\right)}\cos\left(q^ix^i\right),\\
X^{(2)}_i(x^i)&\equiv&\sum_{{q}^i}X^{(2)}_{\left(0,\ldots,q^i,0,\ldots,0\right)}\sin\left(q^ix^i\right),\label{3.38}
\end{eqnarray}
and their respective conjugate momenta
\begin{eqnarray}
P^{(1)}_i(x^i)&\equiv&\sum_{{q}^i}P^{(1)}_{\left(0,\ldots,q^i,0,\ldots,0\right)}\cos\left(q^ix^i\right),\\
P^{(2)}_i(x^i)&\equiv&\sum_{{q}^i}P^{(2)}_{\left(0,\ldots,q^i,0,\ldots,0\right)}\sin\left(q^ix^i\right).\label{3.39}
\end{eqnarray}
In terms of the original variables $S_{\bf r}$ and $\Pi_{\bf r}$, we have
\begin{eqnarray}
X^{(1)}_i(x^i)&=&\frac{1}{2}\sum_{\overset{\{x_j\}}{j\neq i}}\left(S_{\left(x_1,\ldots,x_i,\ldots,x_d\right)}+S_{\left(x_1,\ldots,-x_i,\ldots,x_d\right)}\right),\nonumber\\
P^{(1)}_i(x^i)&=&\frac{1}{2}\sum_{\overset{\{x_j\}}{j\neq i}}\left(\Pi_{\left(x_1,\ldots,x_i,\ldots,x_d\right)}+\Pi_{\left(x_1,\ldots,-x_i,\ldots,x_d\right)}\right),\nonumber\\
X^{(2)}_i(x^i)&=&\frac{1}{2 }\sum_{\overset{\{x_j\}}{j\neq i}}\left(S_{\left(x_1,\ldots,x_i,\ldots,x_d\right)}-S_{\left(x_1,\ldots,-x_i,\ldots,x_d\right)}\right),\nonumber\\
P^{(2)}_i(x^i)&=&\frac{1}{2 }\sum_{\overset{\{x_j\}}{j\neq i}}\left(\Pi_{\left(x_1,\ldots,x_i,\ldots,x_d\right)}-\Pi_{\left(x_1,\ldots,-x_i,\ldots,x_d\right)}\right).\label{3.40}
\end{eqnarray}
We see that $X^{(1)}_i$ and $X^{(2)}_i$ are merely the even and odd parts of the function $X_i(x_i)\equiv\sum_{\overset{\{x_j\}}{j\neq i}}S_{\bf r}$,  whereas $P^{(1)}_i$ and $P^{(2)}_i$ are the even and odd parts of $P_i(x_i)\equiv\sum_{\overset{\{x_j\}}{j\neq i}}\Pi_{\bf r}$, respectively. Therefore, we collect the infinite set of order parameter operators into the quasi-local functions
\begin{equation}
X_i(x_i)=X_i^{(1)}(x_i)+X_i^{(2)}(x_i)=\sum_{\overset{\{x_j\}}{j\neq i}}S_{\bf r},~~~i=1,\ldots,d,\label{3.41}
\end{equation}
with
the respective conjugate momenta
\begin{equation}
P_i(x_i)=P_i^{(1)}(x_i)+P_i^{(2)}(x_i)=\sum_{\overset{\{x_j\}}{j\neq i}}\Pi_{\bf r},~~~i=1,\ldots,d.\label{3.42}
\end{equation}
Also, we  can define the quasi-local magnetization 
\begin{equation}
m_i(x_i)\equiv\sum_{q_i} \left[m^{(1)}_{\left(0,\ldots,q_i,0,\ldots\right)}\cos\left(q_ix_i\right)+m^{(2)}_{\left(0,\ldots,q_i,0,\ldots\right)}\sin\left(q_ix_i\right) \right]
\end{equation}
and construct the ground state as
\begin{eqnarray}
\left|m^{(1)},m^{(2)}\right\rangle_c = \bigotimes_{q^{i}}\bigotimes_{{\bf q}\,\in\,\left[{\bf q}^{c}\right]/\mathbb{Z}_2}|m^{(1)}_{\bf q}\rangle\otimes|m^{(2)}_{\bf q}\rangle.
\end{eqnarray}
Notice that since ${\bf q}^c=\left(0,\ldots,q_i,0,\ldots,0\right)$, after making the product over the set $\left[{\bf q}^{c}\right]/\mathbb{Z}_{2}$, we still have to take into account  all the possible values $q^{i}$. In this way 
\begin{eqnarray}
X_i(x_i)\left|m^{(1)},m^{(2)}\rangle_c\right.&=&\left(X^{(1)}_i(x_i)+X^{(2)}_i(x_i)\right)\bigotimes_{q^{i}}\bigotimes_{{\bf q}\,\in\,\left[{\bf q}^{c}\right]/\mathbb{Z}_2}|m^{(1)}_{\bf q}\rangle\otimes|m^{(2)}_{\bf q}\rangle\nonumber\\
&=&\sum_{q_i}\left(X^{(1)}_{\left(0,\ldots,q^i,0,\ldots,0\right)}\cos\left(q^ix^i\right)+X^{(2)}_{\left(0,\ldots,q^i,0,\ldots,0\right)}\sin\left(q^ix^i\right)\right)|m^{(1)},m^{(2)}\rangle_c\nonumber\\
&=&\sum_{q_i}\left(m^{(1)}_{\left(0,\ldots,q^i,0,\ldots,0\right)}\cos\left(q^ix^i\right)+m^{(2)}_{\left(0,\ldots,q^i,0,\ldots,0\right)}\sin\left(q^ix^i\right)\right)|m^{(1)},m^{(2)}\rangle_c\nonumber\\
&=&m_i(x_i)|m^{(1)},m^{(2)}\rangle_c.
\label{3.43}
\end{eqnarray}
Given arbitrary functions $\alpha(x_i)$, the infinite set of conserved charges $P_i(x_i)$ generates shifts of the fields $S_{\bf r} \rightarrow S_{\bf r}+\alpha(x_i)$ due to the commutation relations 
\begin{equation}
i\sum_{x^\prime_i}\alpha(x^\prime_i)\left[P_i(x_i), S_{\bf r}\right]=\alpha(x_i).
\label{3.44}
\end{equation}
This corresponds to a particular case of the transformation (\ref{0.41c}) involving a single coordinate.

For the case $p>1$ with $j_3=2 j_2$, we have an analogous situation, but for generic $p$ values, the construction of quasi-local order parameter fields may not be convenient. Let us consider the case $d=2$ for simplicity. According to (\ref{0.28}), the set of critical momenta satisfy:
\begin{equation}
\cos(q_2)=\frac{2}{p}-\cos(q_1).\label{3.45}
\end{equation}
For each $p>1$ the possible values of $q_1$ do not cover the entire Brillouin zone, but only the region that obeys $\cos(q_1)<\frac{2}{p}-1$. Then, the value of $p$ defines a positive angle $\alpha=\cos^{-1}\left(1-\frac{2}{p}\right)$ such that $-(\pi-\alpha)<q_1<\pi-\alpha$. For each $q_1$ in this interval, $q_2$ is given by the solution of (\ref{3.45}). Let us denote one of the two solutions of (\ref{3.45}) by $q_2(q_1)$. In this case, we have the set of critical momenta of the type $(q_1,q_2(q_1))$ and $(q_1,-q_2(q_1))$ for all $q_1$ obeying $-(\pi-\alpha)<q_1<\pi-\alpha$. We also have critical momenta interchanging the roles of $q_1$ and $q_2$, i.e., $(q_1(q_2),q_2)$ and $(-q_1(q_2),q_2)$  for all $q_2$ obeying $-(\pi-\alpha)<q_2<\pi-\alpha$.

The main point is that in the regions considered in this section, the structure of the global symmetry group generated by the conserved charges is more general than the class of usual symmetries considered in quantum field theory. The infinite set of conserved charges, for instance, is defined only in submanifolds of the entire space. For the case $p=1$ and $j_2=0$, for instance, the charges $P_i(x_i)$ only act on fields lying on hyperplanes orthogonal to the $x_i$ coordinate. The consequences of the presence of these exotic symmetries are explored in recent investigations in connection with models presenting fractonic degrees of freedom \cite{Fisher,Seiberg1,Seiberg2,Seiberg3}.


\subsection{Translation Symmetry}

We have considered some of the consequences of SSB either when ${\bf q}^c=0$ or ${\bf q}^c\neq 0$. In the first case we only have a $U(1)$ symmetry generated by $P=\Pi_0$ that is spontaneously broken, whereas for ${\bf q}^c\neq 0$ we have at least $U(1)^{\mathsf{N}_c}$ for the broken global symmetry, recalling that $\mathsf{N}_c$ is the number of inequivalent critical momenta obtained with the application of the permutation and sign change operators on a representative critical momentum. Another important difference between the two cases concerns translation symmetry.

So, it is convenient to define $T({\bf a})$ as been the operator that performs translations on the original lattice:
\begin{eqnarray}
T({\bf a})S_{\bf r}T^\dagger({\bf a})&=&S_{{\bf r}+{\bf a}},\label{3.46}\\
T({\bf a})\Pi_{\bf r}T^\dagger({\bf a})&=&\Pi_{{\bf r}+{\bf a}}.\label{3.47}
\end{eqnarray}
From (\ref{3.37}), we see that for $\left[{\bf q}^c\right]\neq 0$ the set of degenerate ground states can be generated from the vacuum with $m^{(1)}=m^{(2)}=0$ by the action of $U(m^{(1)},m^{(2)})$. Therefore, from the translational invariance of $\left|0,0\rangle\right.$, we get
\begin{equation}
T({\bf a})\left|m^{(1)},m^{(2)}\rangle\right.=T({\bf a})U(m^{(1)},m^{(2)})T^\dagger({\bf a})|0,0\rangle.
\label{3.48}
\end{equation}
Using (\ref{3.24}) and (\ref{3.24a}), we have
\begin{eqnarray}
T({\bf a})U(m^{(1)},m^{(2)})T^\dagger({\bf a})&=&\bigotimes_{{\bf q}\,\in\,\left[{\bf q}^c\right]/\mathbb{Z}_2}\exp{i\left(m^{(1)}_{\bf q}\cos\left({\bf q}\cdot{\bf a}\right)-m^{(2)}_{\bf q}\sin\left({\bf q}\cdot{\bf a}\right)\right)P^{(1)}_{\bf q}}\otimes\nonumber\\
&&\bigotimes_{{\bf q}\,\in\,\left[{\bf q}^c\right]/\mathbb{Z}_2}\exp{i\left(m^{(2)}_{\bf q}\cos\left({\bf q}\cdot{\bf a}\right)+m^{(1)}_{\bf q}\sin\left({\bf q}\cdot{\bf a}\right)P^{(2)}_{\bf q}\right)}.
\end{eqnarray}
We conclude that in general the ordered ground state do not share the full translational invariance of the underlying lattice. The general lattice translations are spontaneously broken down to the subgroup $\tilde{T}({\bf a})$, where ${\bf a}$ satisfies ${\bf q}_{i}\cdot{\bf a}=2n_{i}\pi$, with $n_{i}\in\mathbb{Z}$ and ${\bf q}_i$ $\in$ $\left[{\bf q}^c\right]$. Since ${\bf a}$ is in the original lattice, we can only have residual translation invariance if all the ${\bf q}_i$ in $\left[{\bf q}^c\right]$ are commensurate with the reciprocal lattice vectors. The original lattice is hypercubic and we have normalized the lattice spacing to the unity. Therefore, the modes in reciprocal lattice can be written as ${\bf q}=2\pi\left(n_1,\ldots,n_d\right)$, with $n_i \in \mathbb{Z}$, and there will be some residual translational symmetry if ${\bf q}^c=2\pi\left(\frac{n_1}{m_1},\ldots,\frac{n_d}{m_d}\right)$, with both $n_i$ and $m_i$ $\in$ $\mathbb{Z}$.  Therefore, for $d=2$, except in the regions $(I)$ and $(IV)$, and for $d=3$, except in the regions $(I)$ and $(V)$, there are several ordered phases without any kind of translation invariance.

Let us discuss how to define wave excitations propagating on some arbitrary modulated vacuum for a general $\left[{\bf q}^c\right]$, which can break the lattice translation symmetry completely. 

Generically, for $\mu=\mu_c$ we can write the full Hamiltonian as
\begin{equation}
\mathcal{H}=\mathcal{H}_{\left[{\bf q}^c\right]}+\sum_{{\bf q}\,\notin\,\left[{\bf q}^c\right]}\mathcal{H}_{{\bf q}}\label{3.49}
\end{equation}
where
\begin{equation}
\mathcal{H}_{{\bf q}}=\omega_{{\bf q}}\left(a^{\dagger}_{\bf q}a_{\bf q}+\frac{1}{2}\right).
\label{3.50}
\end{equation}
From (\ref{3.46}) and (\ref{3.47}), we see that
\begin{eqnarray}
T({\bf R})a_{\bf q}T^{\dagger}({\bf R})&=&e^{i{\bf q}\cdot{\bf R}}a_{\bf q}\\\label{3.51}
T({\bf R})a^\dagger_{\bf q}T^{\dagger}({\bf R})&=&e^{-i{\bf q}\cdot{\bf R}}a^\dagger_{\bf q}\label{3.52},
\end{eqnarray}
which imply $\left[T({\bf R}),\mathcal{H}_{\bf q}\right]=0$. Therefore, as usual, for each non-critical mode, we construct states $\left|{n_{\bf q}}\rangle\right.$ with $n_{\bf q}$ ${\bf q}$-modes from a given translation invariant vacuum $\left|0\rangle\right.$ by successive application of the creation operators $a^\dagger_{\bf q}$. These states are translational invariant using (\ref{3.52}). For instance,
\begin{equation}
T({\bf R})a^{\dagger}_{\bf q}\left|0\rangle\right.=T({\bf R})\left|{\bf q}\rangle\right.=e^{-i{\bf q}\cdot{\bf R}}\left|{\bf q}\rangle\right..
\label{3.53}
\end{equation}
 For the critical mode Hamiltonian, as we have seen, the ground state $\left|m^{(1)},m^{(2)}\rangle_{\left[{\bf q}^c\right]}\right.$ breaks the symmetry generated by the $\mathsf{N}_c$ charges $P^{(i)}_{{\bf q}}$, with ${\bf q}$ $\in$ $\left[{\bf q}^c\right]/\mathbb{Z}_2$ and $i=1,2$. Since, we have a decoupled set of systems, the full ground state of the model is given by
\begin{equation}
\left|\Omega\rangle_{\left[{\bf q}^c\right],m}\right.=\left|m^{(1)},m^{(2)}\rangle_{\left[{\bf q}^c\right]}\right.\otimes\left|0\rangle\right..\label{3.54}
\end{equation}
The operators $X_{\bf q}$ and $a_{\bf q}$ act on $\left|\Omega\rangle_{\left[{\bf q}^c\right],m}\right.$ as
\begin{eqnarray}
a_{\bf q}\left|\Omega\rangle_{\left[{\bf q}^c\right],m}\right.&=&0,~~~\text{for}~~~{\bf q} \notin \left[{\bf q}^c\right],\\
X^{(i)}_{{\bf q}}\left|\Omega\rangle_{\left[{\bf q}^c\right],m}\right.&=&m^{(i)}_{{\bf q}}\left|\Omega\rangle_{\left[{\bf q}^c\right],m}\right.,~~~\text{for}~~~{\bf q} \in \left[{\bf q}^c\right],  \label{3.55}
\end{eqnarray}
such that we define an excited state $\left|{\bf q}\rangle_{\left[{\bf q}^c\right],m}\right.$ above the ordered ground state $\left|\Omega\rangle_{\left[{\bf q}^c\right],m}\right.$ by
\begin{equation}
\left|{\bf q}\rangle_{\left[{\bf q}^c\right],m}\right.\equiv a^{\dagger}_{\bf q}\left|\Omega\rangle_{\left[{\bf q}^c\right],m}\right.=\left|m^{(1)},m^{(2)}\rangle_{\left[{\bf q}^c\right]}\right.\otimes\left|{\bf q}\rangle\right..
\label{3.56}
\end{equation}
In spite of the biased notation, $\left|{\bf q}\rangle_{\left[{\bf q}^c\right],m}\right.$ is not an eigenstate of the translation operator due to the non-invariance of the critical mode sector $\left|m^{(1)},m^{(2)}\rangle_{\left[{\bf q}^c\right]}\right.$. In fact, from (\ref{3.48}) and (\ref{3.54}), we get
\begin{equation}
T({\bf R})\left|{\bf q}\rangle_{\left[{\bf q}^c\right],m}\right.=e^{-i{\bf q}\cdot{\bf R}}\left|{\bf q}\rangle_{\left[{\bf q}^c\right],\tilde{m}}\right.,\label{3.57}
\end{equation}
which is not an excited state over $\left|\Omega\rangle_{\left[{\bf q}^c\right],m}\right.$, but over the rotated ground state:
\begin{eqnarray}
\left|\Omega\rangle_{\left[{\bf q}^c\right],\tilde{m}}\right.&=&\bigotimes_{{\bf q}\,\in\,\left[{\bf q}^c\right]/\mathbb{Z}_2}\left|m^{(1)}_{\bf q}\cos\left({\bf q}\cdot{\bf R}\right)-m^{(2)}_{\bf q}\sin\left({\bf q}\cdot{\bf R}\right)\rangle\right.\nonumber\\
&&\bigotimes_{{\bf q}\,\in\,\left[{\bf q}^c\right]/\mathbb{Z}_2}\left|m^{(2)}_{\bf q}\cos\left({\bf q}\cdot{\bf R}\right)+m^{(1)}_{\bf q}\sin\left({\bf q}\cdot{\bf R}\right)\rangle\right.. 
\end{eqnarray}
So, what is the meaning of the label ${\bf q}$? To clarify this point, let us represent $T({\bf R})$ in terms of the $a$ and $a^\dagger$ operators:
\begin{equation}
T({\bf R})=\exp\left(i\sum_{{\bf q}}a^{\dagger}_{\bf q}a_{\bf q}\,{\bf q}\cdot{\bf R}\right).
\label{3.58}
\end{equation}
Then, it is clear the way $T({\bf R})$ acts on the full state by translating all the modes. To obtain a unitary operator for which the states $\left|{\bf q}\rangle_{m}\right.$ are eigenstates with well defined quantum number ${\bf q}$, we define the operator that only translates the non-critical modes
\begin{equation}
T_{\left[{\bf q}^c\right]}({\bf R})=\exp\left(i\sum_{{\bf q}\notin\left[{\bf q}^c\right]}a^{\dagger}_{\bf q}a_{\bf q}\,{\bf q}\cdot{\bf R}\right),\label{3.59}
\end{equation}
for which
\begin{equation}
T_{\left[{\bf q}^c\right]}({\bf R})\left|{\bf q}\rangle_{m}\right.=e^{-i{\bf q}\cdot{\bf R}}\left|{\bf q}\rangle_{m}\right..
\end{equation}
In addition, $[T_{\left[{\bf q}^c\right]}({\bf R}),\mathcal{H}]=0$ and then ${\bf q}$ is a good conserved quantum number.
Since $\mathcal{H}_{\left[{\bf q}^c\right]}\left|m^{(1)},m^{(2)}\rangle_{\left[{\bf q}^c\right]}\right.=0$, from (\ref{3.49}), (\ref{3.50}), and (\ref{3.56}), we get
\begin{equation}
\mathcal{H}\left|{\bf q}\rangle_{\left[{\bf q}^c\right],m}\right.=\frac{3}{2}\omega_{{\bf q}}\left|{\bf q}\rangle_{\left[{\bf q}^c\right],m}\right..
\end{equation}
As $\omega_{{\bf q}}\rightarrow 0$ when $\left|{\bf q}\right|\rightarrow \left|{\bf q}^c\right|$, the spectrum is gapless and the modes $\left|{\bf q}\rangle_{m}\right.$ are the Nambu-Goldstone excitations for the spontaneous breaking of the global symmetries.


\subsection{Counting the Nambu-Goldstone Bosons}

Let us discuss more precisely the counting of Goldstone bosons in all the gapless phases. According to the previous section, we get only one Godstone excitation in the spectrum regardless if ${\bf q}^c=0$ or ${\bf q}^c\neq 0$. This seems a little surprising due to the large group of global symmetry $U(1)^{\mathsf{N}_c}$ that is spontaneously broken for ${\bf q}^c\neq 0$. In the modern scheme \cite{Watanabe1,Hidaka,Watanabe2,Watanabe3}, the Goldstone excitations due to the breaking of uniform symmetries can be classified in two types according to the form of the most relevant operator involving time derivatives appearing in the low-energy effective action. For an operator with two time derivatives, they are referred to as {\it type-A}. This case encompasses the relativistic setting and the relation between the number of NG bosons and broken generators is one-to-one. On the other hand, when the most relevant operator in the effective action involves a single time derivative, pairs of canonical momenta of the Goldstone fields get constrained to be conjugate to each other, therefore reducing the number of independent excitations. These Goldstone excitations are called {\it type-B}. The general counting rule turns out to be \cite{Watanabe1,Hidaka,Watanabe2,Watanabe3}
\begin{equation}
n_A= n_{BG}-\text{rank}(\rho)~~~\text{and}~~~n_B=\frac{1}{2} \text{rank}(\rho),\label{7.1}
\end{equation}
where $n_A$ and $n_B$ are the number of type-A and type-B NG bosons, $n_{BG}$ is the number of broken generators, and $\rho$ is a matrix calculated from the broken charge generators $Q_i$ as  
\begin{equation}
\rho_{ij}\equiv-\frac{i}{V} \langle [Q_i,Q_j]\rangle.\label{7.2}
\end{equation} 
In simple scenarios the symmetry generators commute among themselves so that $\text{rank}(\rho)=0$, and we end up with the most common situation where the number of NG is the same as the number of broken symmetry generators. For this case, the NG excitations are all of type-A.

To place our discussion in this classification, we notice that in the lattice Hamiltonian (\ref{4.24}) the kinetic term has two time derivatives. Then, it would be tempted to classify our NG excitations as type-A modes. Besides, in our case, the broken charge generators $P^{(i)}_{\bf q}$ satisfy $[P^{(1)}_{{\bf q}},P^{(2)}_{{\bf q}'}]=0$, which gives $\text{rank}(\rho)=0$ leading to type-A modes. However, this  classification only applies when the symmetry is uniform, i.e., the broken charge densities do not have explicit spacetime coordinates dependence \cite{Watanabe}. For the case with ${\bf q}^c=0$ the gapless phase is homogeneous and the broken charge density $\phi({\bf r})=\Pi_{\bf r}$ satisfies the uniform criterion, i.e., it only has implicit spacetime dependence. This, in turn, can be expressed in terms of the behavior of the charge density under translations,
\begin{equation}
T({\bf a})\phi({\bf r})T^\dagger({\bf a})=\phi({\bf r}+{\bf a}).\label{7.3}
\end{equation}
Therefore, in the case of homogeneous ordered phase ${\bf q}^c=0$, we have one type-A Goldstone boson associated with the broken charge density $\mathcal{P}({\bf r})$, and the counting agree with the general counting rule (\ref{7.1}), giving $n_A=n_{BG}=1$.

For the cases of modulated ordered phases, ${\bf q}^c\neq 0$, the situation is more subtle. In these cases the charge densities do have explicit spatial dependence and the symmetry is classified as non-uniform, so that the counting (\ref{7.1}) does not apply. 

When ${\bf q}^c$ is of the form ${\bf q}^c=\left(n_1\pi,\ldots,n_d\pi\right)$ with $n_i=0,1$, the charge densities can be written as $\phi_{\bf q}({\bf r})=\Pi_{\bf r}(-1)^{n_1x_1+\ldots+n_dx_d}$, which do not satisfy the transformation property (\ref{7.3}). For two distinct critical momenta ${\bf q}=\left(n_1\pi,\ldots,n_d\pi\right)$ and ${\bf q}'=\left(n'_1\pi,\ldots,n'_d\pi\right)$ $\in$ $\left[{\bf q}^c\right]$, we can relate the broken charge densities through $\phi_{\bf q}({\bf r})=\phi_{\bf q'}({\bf r})(-1)^{(n_1-n'_1)x_1+\ldots+(n_d-n'_d)x_d}$. Since waves created in the ordered phases by the action of the local symmetry generators are NG excitations, any $\phi_{\bf q}({\bf r})$ creates the same NG mode when applied on this modulated ordered vacuum. Consequently, the spontaneous breaking of the $U(1)^{\mathsf{N}_c}$ symmetry group gives rise to a single Goldstone boson. 

For the remaining cases with ${\bf q}^c\neq 0$ (including the exotic ones), we have two charge densities, $\phi^{(1)}_{{\bf q}}({\bf r})\equiv\Pi_{\bf r}\cos\left({\bf q}\cdot{\bf r}\right)$ and $\phi^{(2)}_{{\bf q}}({\bf r})\equiv\Pi_{\bf r}\sin\left({\bf q}\cdot{\bf r}\right)$, for each ${\bf q}$ $\in$ $\left[{\bf q}^c\right]/\mathbb{Z}_2$. Again, the explicit spatial dependence of the densities leads to the spoiling of the simple transformation (\ref{7.3}). As in the previous case, we can relate all the charge densities. To this, it is convenient to define the complex fields $\phi_{\bf q} ({\bf r})\equiv\phi^{(1)}_{{\bf q}}+i\phi^{(2)}_{{\bf q}}=\Pi_{\bf r}e^{i{\bf q}\cdot{\bf r}}$ and $\phi^\dagger_{\bf q} ({\bf r})\equiv\phi^{(1)}_{{\bf q}}-i\phi^{(2)}_{{\bf q}}=\Pi_{\bf r}e^{-i{\bf q}\cdot{\bf r}}$, which implies $\phi_{\bf q} ({\bf r})=e^{2i{\bf q}\cdot{\bf r}}\phi^\dagger_{\bf q} ({\bf r})$. Therefore, $\phi_{\bf q} ({\bf r})$ and $\phi^\dagger_{\bf q} ({\bf r})$ create the same excitation. Furthermore, we can also relate the fields $\phi_{\bf q} ({\bf r})$ and $\phi_{{\bf q}'} ({\bf r})$ for distinct ${\bf q}$ and ${\bf q}'$ through $\phi_{{\bf q}'} ({\bf r})=e^{i\left({\bf q}-{\bf q}'\right)\cdot{\bf r}}\phi_{\bf q} ({\bf r})$. Therefore, for any critical mode, the spontaneous breaking of the $U(1)^{\mathsf{N}_c/2}\times U(1)^{\mathsf{N}_c/2}$ symmetry group gives rise to only one Goldstone boson, like in the previous case.

To summarize, in both cases, ${\bf q}^c\neq 0$ and ${\bf q}^c=0$,  there is only one NG boson in the spectrum of the ordered phases in spite of the fact that the global symmetries that are spontaneously broken are significantly different in the respective cases.

\begin{figure}[h!]
	\centering
	\subfloat[\label{correlation1-(a)}]{ %
		\includegraphics[width=0.48\columnwidth]{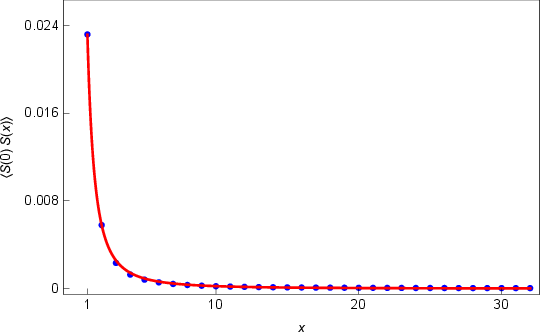}%
	}\hfill
	\subfloat[\label{correlation1-(b)}]{ %
		\includegraphics[width=0.48\columnwidth]{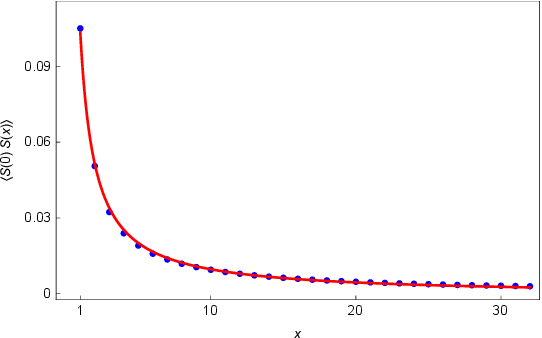} %
	} \hfill
	
	\subfloat[\label{correlation1-(c)}]{ %
		\includegraphics[width=0.48\columnwidth]{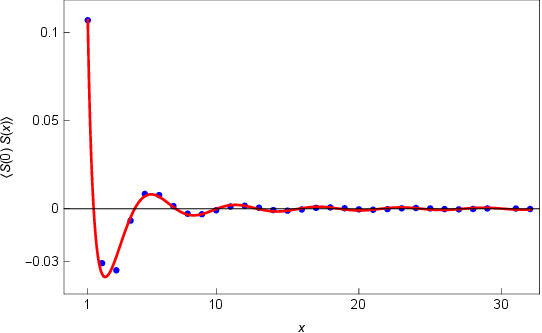} %
	} \hfill
	
	\caption{In these diagrams we show the behavior of the correlations $\langle S_{\bf r}S_{{\bf r}\prime}\rangle$ in the ordered phases for $d=3$ with $j_{3} < 2 j_{2}$ . Figures \protect\subref{correlation1-(a)}, \protect\subref{correlation1-(b)} and \protect\subref{correlation1-(c)} show the evolution of these correlations as we go from region (I) (figure \protect\subref{correlation1-(a)}), crossing the line $p=1$ (figure \protect\subref{correlation1-(b)}), and ending up in region $II$ (figure \protect\subref{correlation1-(c)}) of the diagram \ref{diagram3d}. The correlations were calculated along the directions $x$ or $y$ and in all cases we used $j_{2} = 0.1$, $j_{3} = 0.11$, and $j_{1}$ was set in each case to satisfy $p=0.5$, $p=1$, and $p=2$ for \protect\subref{correlation1-(a)}, \protect\subref{correlation1-(b)}, and \protect\subref{correlation1-(c)}, respectively.}
	\label{correlations1}
\end{figure}

\begin{figure}[h!]
	\subfloat[\label{correlation2-(a)}]{ %
		\includegraphics[width=0.48\columnwidth]{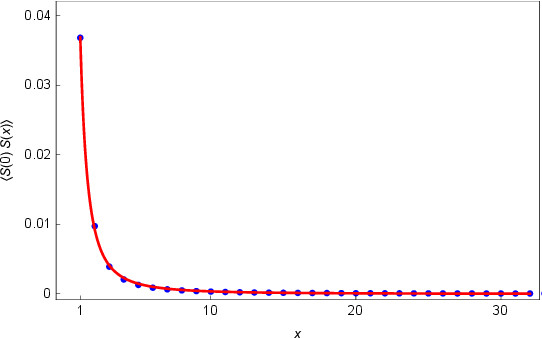} %
	}\hfill 
	\subfloat[\label{correlation2-(b)}]{ %
		\includegraphics[width=0.48\columnwidth]{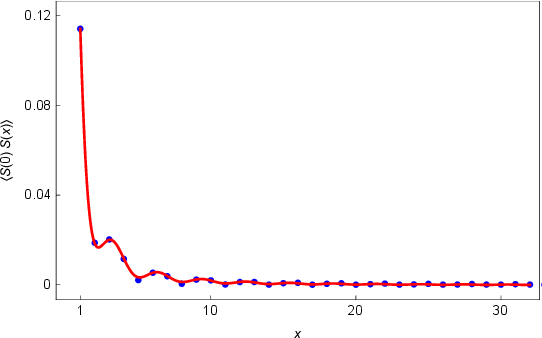} %
	}\hfill 
	\subfloat[\label{correlation2-(c)}]{ %
		\includegraphics[width=0.48\columnwidth]{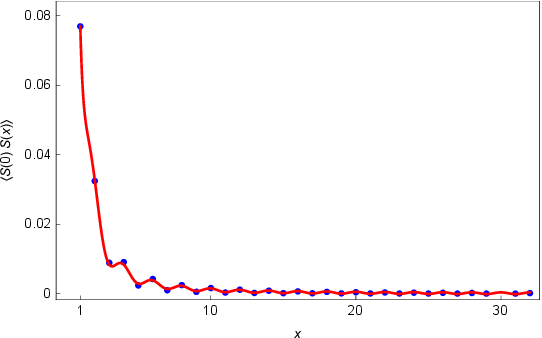} %
	}\hfill 
	\subfloat[\label{correlation2-(d)}]{ %
		\includegraphics[width=0.48\columnwidth]{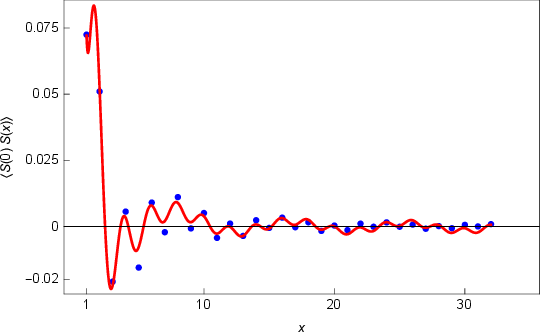} %
	}\hfill 
	\caption{In these diagrams we show the behavior of the correlations $\langle S_{\bf r}S_{{\bf r}\prime}\rangle$ in the ordered phases for $d=3$ with $j_{3}>2 j_{2}$ . Figures \protect\subref{correlation2-(a)}-\protect\subref{correlation2-(d)} correspond to the sequence of regions $I$, $III$, $IV$, and $V$ shown in Fig. \ref{diagram3d}.	These correlations were calculated along one of the three orthogonal directions $x$, $y$, or $z$ and in all cases we used $j_{2} = 0.1$ and $j_{3} = 0.25$ and $j_{1}$ was set in each case to satisfy $p=0.5$, $p=1.7$, $p=2.9$ and $p=5$ for \protect\subref{correlation2-(a)}, \protect\subref{correlation2-(b)}, \protect\subref{correlation2-(c)} and \protect\subref{correlation2-(d)}, respectively.}
	\label{correlations2}
\end{figure}

We finish our discussion by probing numerically the propagation of a massless excitation in the ordered and modulated phases. In Figures (\ref{correlations1}) and (\ref{correlations2}) we show some diagrams of the correlations $\langle S_{\bf r}S_{{\bf r}^\prime}\rangle$ calculated in these phases for $d=3$. All of them have an enveloping curve that falls as a power law $\sim 1/x^2$ (except at the Lifshitz point $p=1$, where the power law goes as $\sim 1/x$), as expected for a gapless phase in $d=3$. Besides, we notice the role of the non-homogeneous ground stated in causing a modulation of the correlations. The sequence shows how the correlations behave as some regions of diagram \ref{diagram3d} are crossed. 

 \newpage
 
 ${}$
 
 \newpage

\section{Final Remarks\label{sec7}}

The phase structure generated by the frustrated interactions in the quantum spherical model is quite remarkable. It exhibits a gapped-to-gapless phase transition corresponding to the transition from a phase where the $U(1)^{\mathsf{N}_c}$ symmetry is explicitly broken (gapped phase) to a phase where the symmetry is spontaneously broken (gapless phase). In addition, there are many gapless-to-gapless continuous quantum phase transitions between different modulated phases. 
We have described the mechanism of phase transitions and we completely characterized the homogeneous and modulated ordered phases by means of the analysis of the spontaneously broken internal and space-time symmetry groups, with the construction of the exact stable ground state in each phase. The only excitations in these phases are NG bosons. Although there may be many broken continuous symmetries (possibly $\mathsf{N}_c$), we always find a single Goldstone excitation because of the non-uniform nature of the broken global symmetries.

The studies pursued here naturally lead us to many questions and also point out to new directions. The very first question is to understand to what extent the peculiar properties presented by the frustrated QSM, like the gapless-to-gapless phase transitions, are model-dependent. In this sense, it would be helpful to examine this particular setting of competing interactions in different systems with the potential to exhibit Goldstone excitations, possibly involving $SU(2)$ Heisenberg spins. 

Concerning the QSM itself, we have shown that the model is able to order in $d=3$ when $p=1$, with $j_2=0$. In this case, the phase supports Goldstone modes with a dispersion relation of the type $E^2\sim q_x^2q_y^2+q_x^2q_z^2+q_y^2q_z^2$. This, in turn, may be associated with an exotic form of symmetry $S_{\bf r}\rightarrow S_{\bf r}+ f_1(x)+f_2(y)+f_3(z)$, which has been recently identified  in certain fracton field theories \cite{Fisher,Seiberg1,Seiberg2,Seiberg3}. It would be interesting to contrast this result with one of Ref. \cite{Seiberg2}, where the authors show that in the case of a {\it compact} scalar field, this symmetry that appears classically (spontaneously) broken is actually restored in the quantum theory. We intend to address this issue in future investigations.

Finally, it would be enlightening to investigate how these properties manifest in a continuum quantum field theory. To address this point, we could use the well known connection between the QSM and the $O(N)$ nonlinear sigma model (NLSM) in the large $N$ limit. In this way, by further exploring this equivalence for the QSM with frustrated interactions would provide a NLSM with the peculiar properties mentioned right above. Furthermore, even though the underlying lattice breaks continuous rotation and translation invariance, these symmetries may emerge at low energies in the continuum limit. The spontaneous breaking of these emergent symmetries can give rise to quasi-NG bosons and the relation between the number of gapless excitations with the number of broken global charges should be readdressed. 

\section{Acknowledgments}

We acknowledge the financial support from the Brazilian funding agencies CAPES and CNPq.



\end{document}